\documentclass{ecai}
\usepackage{times}
\usepackage{graphicx}
\usepackage{latexsym}

\usepackage{wrapfig}

\usepackage{mathtools}
\usepackage{graphicx}
\usepackage{courier}
\usepackage{epstopdf}
\usepackage{color}
\usepackage{csquotes}

\usepackage{amsmath}
\usepackage{amsfonts}
\usepackage{amssymb}
\usepackage[subnum]{cases}
\usepackage{booktabs} %For formal tables
\usepackage{pifont}
\usepackage{graphics}

\usepackage{amssymb}% http://ctan.org/pkg/amssymb
\usepackage{pifont}% http://ctan.org/pkg/pifont

\graphicspath{ {./Images/} }

\usepackage {tikz}
\usetikzlibrary {positioning}
\definecolor {processblue}{cmyk}{0.96,0,0,0}
\usepackage{lipsum,adjustbox}

\usepackage[subnum]{cases}
\usepackage{color,soul}

\usepackage{algorithm,algorithmicx,algpseudocode}

\usepackage{subcaption}

\usepackage{relsize}

\newtheorem{lemma}{Lemma}
\newtheorem{proof}{Proof}

\usepackage{footnote}
\makesavenoteenv{tabular}

\usepackage{multirow}

\usepackage{hyperref}
\usepackage{cleveref}[2012/02/15]% v0.18.4; 
\crefformat{footnote}{#2\footnotemark[#1]#3}

\begin{document}

\title{Beyond Node Embedding: A Direct Unsupervised Edge Representation Framework for Homogeneous Networks}

\author{Sambaran Bandyopadhyay\institute{IBM Research \& IISc, Bangalore, email: sambband@in.ibm.com} \and Anirban Biswas\institute{Indian Institute of Science, Bangalore, email: anirbanb@iisc.ac.in} \and Narasimha Murty\institute{Indian Institute of Science, Bangalore, email: mnm@iisc.ac.in} \and Ramasuri Narayanam\institute{IBM Research, Bangalore, email: ramasurn@in.ibm.com}}

\maketitle
\bibliographystyle{ecai}

\begin{abstract}
Network representation learning has traditionally been used to find lower dimensional vector representations of the nodes in a network. However, there are very important edge driven mining tasks of interest to the classical network analysis community, which have mostly been unexplored in the network embedding space. For applications such as link prediction in homogeneous networks, vector representation (i.e., embedding) of an edge is derived heuristically just by using simple aggregations of the embeddings of the end vertices of the edge. Clearly, this method of deriving edge embedding is suboptimal and there is a need for a dedicated unsupervised approach for embedding edges by leveraging edge properties of the network.

Towards this end, we propose a novel concept of converting a network to its weighted line graph which is ideally suited to find the embedding of edges of the original network. We further derive a novel algorithm to embed the line graph, by introducing the concept of \textit{collective homophily}. To the best of our knowledge, this is the \textit{first direct unsupervised approach for edge embedding in homogeneous information networks}, without relying on the node embeddings. We validate the edge embeddings on three downstream edge mining tasks. Our proposed optimization framework for edge embedding also generates a set of node embeddings, which are not just the aggregation of edges. Further experimental analysis shows the connection of our framework to the concept of node centrality.
\end{abstract}

\section{Introduction}\label{sec:intro}
Network representation learning (also known as network embedding) has gained significant interest over the last few years. Traditionally, network embedding \cite{perozzi2014deepwalk,grover2016node2vec,velivckovic2017graph} maps the nodes of a homogeneous network (where nodes denote entities of similar type) to lower dimensional vectors, which can be used to represent the nodes. It has been shown that such continuous node representations outperform conventional graph algorithms \cite{adamic2003friends} on several node based downstream mining tasks like node classification, community detection, etc. 

Edges are also important components of a network. From the point of downstream network mining analytics, there are plenty of network applications - such as computing edge betweenness centrality \cite{Newman:2010} and information diffusion \cite{Rogers:95} - which heavily depend on the information flow in the network. 
Compared to the conventional downstream node embedding tasks (such as node classification), these tasks are more complex in nature. But similar to node based analytics, there is a high chance to improve the performance of these tasks in a continuous lower dimensional vector space. Thus, it makes sense to address these problems in the context of network embedding via direct representation of the edges of a network. As a first step towards this direction, it is important to design dedicated edge embedding schemes and validate the quality of those embeddings on some basic edge-centric downstream tasks. 

\begin{figure}[h!]
  \centering
  \begin{subfigure}[b]{0.32\linewidth}
    \includegraphics[width=\linewidth]{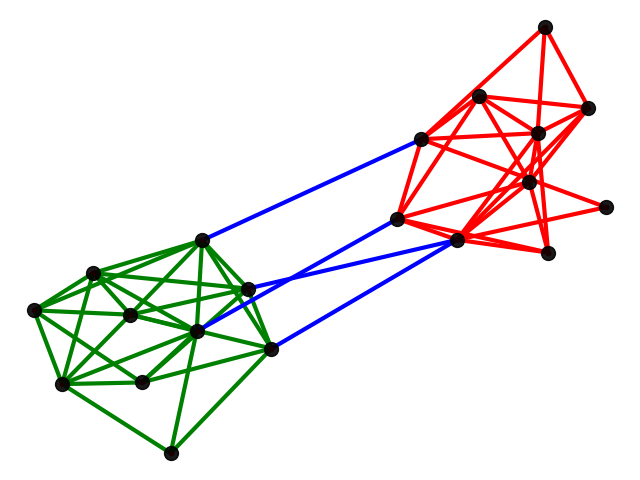}
    \caption{Synthetic Graph}
  \end{subfigure}
  \begin{subfigure}[b]{0.32\linewidth}
    \includegraphics[width=\linewidth]{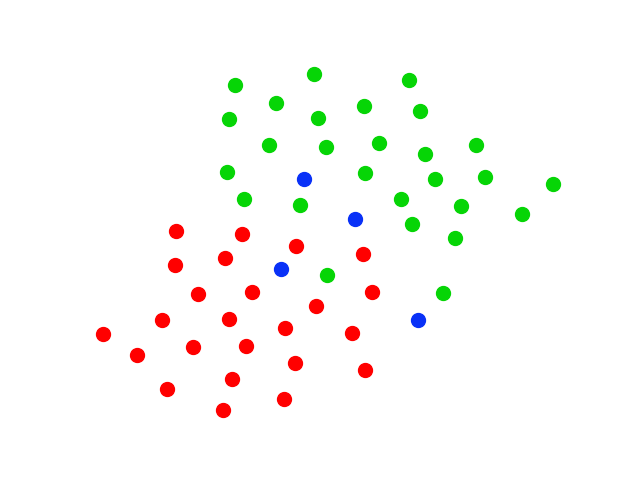}
    \caption{node2vec}
  \end{subfigure}
  \begin{subfigure}[b]{0.32\linewidth}
    \includegraphics[width=\linewidth]{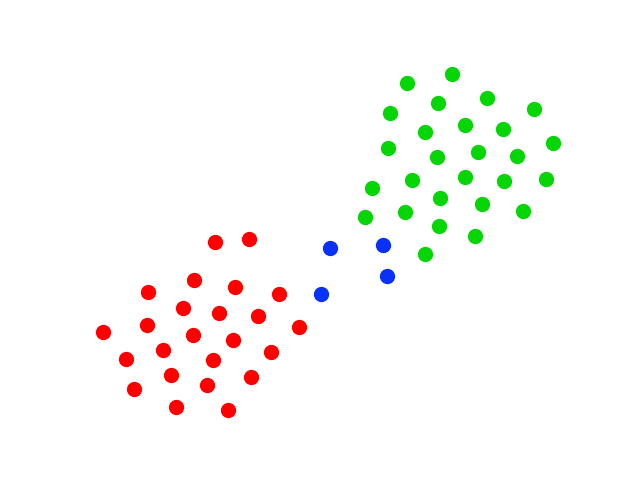}
    \caption{line2vec}
  \end{subfigure}
  \caption{Edge Visualization: (a) We created a small synthetic network with two communities. So, there are three types of edges: Green (or red) edges with both the end points belonging to the green (or red respectively) community; Blue edges with end points belonging to two different communities. (b) node2vec embedding (8 dimensional) of the edges obtained by taking average of the embeddings of the end vertices and then used t-SNE for visualization. (c) Direct edge embeddings (8 dimensional) obtained by line2vec and then used t-SNE for visualization. Clearly, line2vec is superior which visually separates the edge communities, compared to that with the conventional way of aggregating node embeddings to obtain edge representation.}
  \label{fig:motiv}
\end{figure}

In the literature, there are indirect ways to compute embedding of an edge in an information network. For tasks like link prediction, where a classifier needs to be trained on both positive (existing) and negative (not existing) edge representations, a simple aggregation function \cite{grover2016node2vec} such as vector average or Hadamard product has been used on the representations of the two end vertices to derive the vector representation of the corresponding edge. 
Typically node embedding algorithms use the homophily property \cite{mcpherson2001birds} by respecting different orders of node proximities in a network.
As the inherent objective functions of these algorithms are focused on the nodes of the network, using an aggregation function on these node embeddings to get the edge embedding could be  suboptimal.
We demonstrate the shortcoming of this approach in Figure \ref{fig:motiv}, where the visualization of the edge embeddings derived by aggregating node embeddings (taking average of the two end nodes) from node2vec \cite{grover2016node2vec} on a small synthetic graph do not maintain the edge community structure of the network. Whereas, a direct edge embedding approach line2vec, to be proposed in this paper, completely adheres to the community structure, as edges of different types are visually segregated in the t-SNE plot of the same shown in \ref{fig:motiv}(c). So there is a need to develop algorithms for directly embedding edges (i.e., not via aggregating node embeddings) in information networks. We address this research gap in this paper in a natural way. Following are the contributions:% we make:
\begin{itemize}
\item We propose a novel edge embedding framework \textbf{line2vec}, for homogeneous social and information networks. To the best of our knowledge, this is the first work to propose a {\em dedicated unsupervised edge embedding scheme which avoids aggregation of the end node embeddings}.
\item We exploit the concept of \textit{line graph} for edge representation by converting the given network to a \textit{weighted} line graph. We further introduce the concept of \textit{collective homophily} to embed the line graph and produce the embedding of the edges of the given network.
\item We conduct experiments on three edge-centric downstream tasks. Though our approach is proposed for embedding edges, we further analyze to show that, a set of robust node embeddings, which are not just the aggregation of edges, are also generated in the process.
\item We experimentally discover the non-trivial connection of the classical concept of node centrality with the optimization framework of line2vec. The source code of line2vec is available at \url{https://bit.ly/2kfiS2l} to ease the reproducibility of the results.
\end{itemize}

Though edge centric network mining tasks such as edge centrality, network diffusion and link prediction can be benefited from edge embeddings, applications of edge embeddings to tackle them is non-trivial and needs a separate body of work. For example, finding central edges in the network amounts to detecting a subset of points in the embedding space which are diverse between each other and represent a majority of the other points. We leave them to be addressed in some future work.

\section{Related Work and Research Gaps}\label{sec:related}
Node embedding in information network has received great interest from the research community. 
We refer the readers to the survey articles \cite{wu2019comprehensive} for a comprehensive survey on network embedding and cite only some of the more prominent works in this paragraph.
DeepWalk \cite{perozzi2014deepwalk} and node2vec \cite{grover2016node2vec} are two node embedding approaches which employ different types of random walks to capture the local neighborhood of a node and maximize the likelihood of the node context.
Struc2vec \cite{ribeiro2017struc2vec} is another random walk based strategy which finds similar embeddings for nodes which are structurally similar.
A deep autoencoder based node embedding technique (SDNE) that preserves structural proximity is proposed in \cite{wang2016structural}.
Different types of node embedding approaches for attributed networks are also present in the literature \cite{yang2015network,bandyopadhyay2018fscnmf,gao2018deep}.
A semi-supervised graph convolution network based node embedding approach is proposed in \cite{kipf2016semi} and further extended in GraphSAGE \cite{hamilton2017inductive} which learns the node embeddings with different types of neighborhood aggregation methods on attributes.
%The work proposed in \cite{xu:2018} consider the aspect of different links having different semantic meanings while generating the node embeddings.
Recently, node embedding based on 
%generative adversarial learning \cite{DBLP:conf/aaai/WangWWZZZXG18}, 
semi-supervised attention networks \cite{velivckovic2017graph}, maximizing mutual information \cite{velivckovic2018deep}, and in the presence of outliers \cite{bandyopadhyay2019outlier} are proposed.% in the literature.
%Moreover, vector representation of a full graph \cite{ying2018hierarchical,morris2019weisfeiler} for tasks like graph classification has also drawn interest among the researchers.

Compared to the above, representing edges in information networks is significantly less matured. 
%Many of the above works use post-processing to derive embedding of an edge just by taking a simple aggregation (such as average) of the two end point node embeddings.
Some preliminary works exist which use random walk on edges for community detection in networks \cite{li2017edge} or to classify large-scale documents into large-scale hierarchically-structured categories \cite{golam2017edge2vec}.
\cite{abu2017learning} focuses on the asymmetric behavior of the edges in a directed graph for deriving node embeddings, but it represents a potential edge just by a scalar which determines its chance of existence. \cite{shi2018easing,verma2019heterogeneous} derive embeddings for different types of edges in a heterogeneous network, but their proposed method essentially uses an aggregation function inside the optimization framework to generate edge embeddings from the node embeddings. For knowledge bases, embedding entities and relation types in a low dimensional continuous vector space \cite{bordes2013translating,chen2019embedding,gao2019edge2vec} have been shown to be useful. But, several fundamental concepts of graph embedding, such as homophily, are not directly applicable to them. 
\cite{monti2018dual} proposes a dual-primal GCN based semi-supervised node embedding approach which first aggregates edge features by convolution, and then learns the node embeddings by employing a graph attention on the incident edge features of a node.
To the best of our knowledge, \cite{zhou2018density} is the only work which proposes a supervised approach based on adversarial training and an auto-encoder, purely for edge representation learning in homogeneous networks. But their framework needs a large amount of labelled edges to train the GAN, which makes it restrictive for real world applications. Hence in this paper, we propose a task-independent unsupervised dedicated edge embedding framework for homogeneous information networks to address the research gaps.

%{\bf Research Gap:} Surprisingly, all the network embedding works mentioned above are for embedding nodes of the network. There is no dedicated scheme for edge embedding, which avoids aggregation of the node embeddings. We highlight that inducing edge embeddings from node embeddings is not effective as we will show in the later part of this paper. Thus, there is a genuine need for direct embedding of the edges of a network and we precisely address this research gap in this paper.

\section{Problem Description}\label{sec:prob}
An information network is typically represented by a graph $G = (V, E, W)$, where $V=\{v_1, v_2,\cdots, v_n\}$ is the set of nodes (a.k.a. vertices), each representing a data object. $E \subseteq \{(v_i,v_j) | v_i,v_j \in V \}$ is the set of edges. We assume, $|E|=m$. Each edge $e \in E$ is associated with a weight $w_{v_i,v_j} > 0$ (1 if $G$ is unweighted), which indicates the strength of the relation. Degree of a node $v$ is denoted as $d_v$, which is the sum of weights of the incident edges. $\mathcal{N}(v)$ is the set of neighbors of the node $v \in V$.
For the given network $G$, the edge representation learning is to learn a function $f : e \mapsto \mathbf{x} \in \mathbb{R}^K$, i.e., it maps every edge $e \in E$ to a $K$ dimensional vector called edge embedding, where $K < m$. These edge embeddings should preserve the underlying edge semantics of the network, as described below.

\textbf{Edge Importance}: Not all the edges in a network are equally important. For example, in a social network, millions of fans can be connected to a movie star. But any two fans of a movie star may not be similar to each other. So this type of connections are weaker compared to an edge which connects two friends who have much lesser number of connections individually \cite{liben2007link}.

\textbf{Edge Proximity}: The edges which are close to each other in terms of their topography or semantics should have similar embeddings. Similar to the concepts of node proximities \cite{wang2016structural}, it is easy to define first and higher order edge proximities via incidence matrix.

%The representation should also be compact and continuous as that would help the downstream machine learning algorithms perform better. 

\section{Solution Approach: line2vec}\label{sec:sol}
We propose an elegant solution (referred as line2vec) to embed each edge of the given network. First we map the network to a weighted line graph, where each edge of the original network is transformed into a node.% and the edge adjacency in the original graph is preserved using the edges in the line graph. 
Then we propose a novel approach for embedding the nodes of the line graph, which essentially provides the edge embeddings of the original network. For simplicity of presentation, we assume that the given network is undirected. Nevertheless, it can trivially be generalized for directed graphs.

\subsection{Line Graph Transformation}\label{sec:LT}
\begin{figure}%[h]
\centering
\includegraphics[scale=0.25]{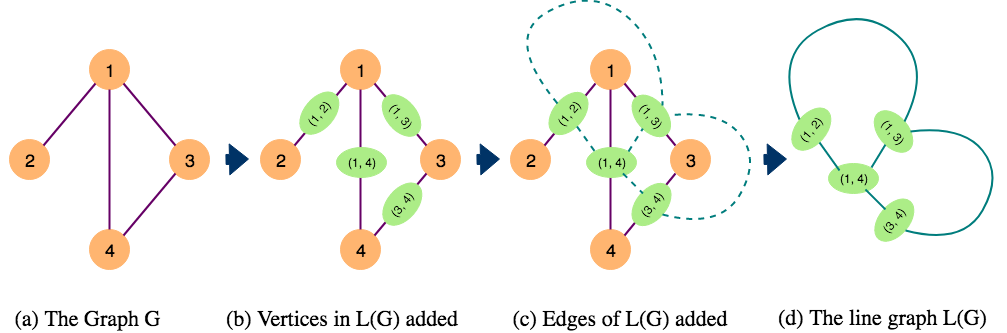}
\caption{Transformation process of a graph into its line graph.
(a) Represents an information network $G$. (b) Each edge in the original graph has a corresponding node in the line graph. Here the green edges represent the nodes in line graph. (c) For each adjacent pair of edges in $G$ there exists an edge in $L(G)$. The dotted lines here are the edges in the line graph. (d) The line graph $L(G)$ of the graph $G$}
\label{fig:line_graph}
\end{figure}

Given an undirected graph $G = (V, E)$, the line graph $L(G)$ is the graph such that each node of $L(G)$ is an edge in $G$ and two nodes of $L(G)$ are neighbors if and only if their corresponding edges in $G$ share a common endpoint vertex \cite{whitney}. Formally $L(G) = (V_{L}, E_{L})$ where $V_L = \{(v_i, v_j) : (v_i, v_j) \in E\}$ and $E_L = \{\big((v_i, v_j), (v_j, v_k)\big) : (v_i, v_j) \in E \:, \: (v_j, v_k) \in E \}$.
Figure \ref{fig:line_graph} shows how to convert a graph into the line graph \cite{evans2010line}.
%The definition of line graph can easily be generalized for directed graph as well. 
%The directed line graph (also called \textit{line digraph}) is defined in a manner similar to its undirected variant. For a directed graph $G_{d} = (V_{d}, E_{d})$, the corresponding line graph $L(G_d) = (V_L^d, E_L^d)$ where $V_L^d = \{(v_i, v_j) : (v_i, v_j) \in E_d\}$ and $E_L^d = \{\big((v_i, v_j), (v_j, v_k)\big) : (v_i, v_j) \in E \:, \: (v_j, v_k) \in E \}$.
Hence the line graph transformation induces a bijection from the set of edges of the given graph to the set of nodes of the line graph as $l: e \mapsto \mathbf{v}$ where $\forall e \in E, \: \exists \: \mathbf{v} \in V_L$ and if two edges $e_i, e_j \in E$ are adjacent there will be an corresponding edge $\mathbf{e} \in E_L$ in the line graph. %The same argument is valid for directed graph also.

\subsection{Weighted Line Graph Formation}\label{sec:weightLG}
We propose to construct a weighted line graph for our problem even if the original graph is unweighted. These weights would help the random walk in the later stage of line2vec to focus more on the relevant nodes in the line graph. It is evident from Section \ref{sec:LT} that a node of degree $k$ in the original graph $G$ produces $k(k-1)/2$ edges in the line graph $L(G)$. Therefore high degree nodes in the original graph may get over-represented in the line graph.
%This type of over-representation may not be good for the analysis of the network. 
Often many of these incident edges are not that important to the concerned node in the given network, but they can potentially change the movement frequency of a random walk in the line graph.
%If the random walk (explained in Section \ref{sec:embedLG}) on the nodes of $G$ passes through an edge $e\in E$ with a frequency $f$ and $e$ is mapped to a vertex $v$ in $L(G)$ then the frequency with which $v$ is visited while doing a random walk on $L(G)$ can be very different. 
%{\color{red}If the degrees of the endpoints of an edge $e \in E$ are $\mathcal{O}{(d)}$ then the corresponding node $\mathbf{v}=l(e)$ in $L(G)$ will be visited $\mathcal{O}{(d^2)}$ times.} 
%
We follow a simple strategy to overcome this problem. %If two edges $e_1$ and $e_2$ in the original undirected graph have a common vertex $v$ with degree $d_v( \geq 2)$ the weight of the edge $(e_1, e_2)$ in the line graph is changed to $\frac{1}{d_v - 1}$. The extension of this strategy for directed as well as weighted graphs is also straightforward as explained later. 
The goal is to ensure that the line graph not only reflects the topology of the original graph $G$ (which is guaranteed by Whitney graph isomorphism theorem \cite{whitney} in almost all cases) but also the dynamics of the graph is not affected by the transformation process.
The edge weights are defined to facilitate a random walk on $L(G)$, as described in Section \ref{sec:collecHom}. Intuitively if we start a random walk from a node $\mathbf{v}_{ij} \equiv (v_i, v_j) \in L(G)$ and want to traverse to $\mathbf{v}_{jk} \equiv (v_j, v_k) \in L(G)$, then it is equivalent to selecting the node $v_j \in G$ from $(v_i, v_j)$ and move to $v_k \in G$. If $G$ is undirected, we define the probability of choosing $v_j$ to be proportional to $\frac{d_{v_i}}{d_{v_i} + d_{v_j}}$. Here, $d_{v_i}$ and $d_{v_j}$ are the degrees of the end point nodes of the edge $(v_i, v_j)$ and an edge in general is more important to the endpoint node having lower degree than the other endpoint with a higher degree \cite{liben2007link}. Then selecting $v_k$ is proportional to edge weight of $e_{jk} \equiv (v_j, v_k) \in E$. 
%In case of a directed network, probability of choosing $v_j$ is 1 as $e_{ij} \equiv (v_i, v_j) \in E$ is a directed edge and selection of node $v_k$ is proportional to the outgoing edge weight $w_{jk}$.
%
Hence,
%when the input network is a weighted undirected graph $G$, 
for any two adjacent edges $e_{ij} \equiv (v_i,v_j)$ and $e_{jk} \equiv (v_j,v_k)$, we define the edge weight for the edge $(e_{ij},e_{jk})$ of the line graph $L(G)$ as follows:
\begin{align}
    \mathbf{w}_{(e_{ij},e_{jk})} = \frac{d_i}{d_i + d_j} \times \frac{w_{jk}} {\sum\limits_{r \in \mathcal{N}(v_j)}{w_{jr}} - w_{ij}}
\end{align}
% When the given network is a weighted directed graph, we define an edge weight of the line graph $L(G)$ as follows:
% \begin{align}
%     \mathbf{w}_{(e_{ij},e_{jk})} = \frac{w_{jk}} {\sum\limits_{r \in \mathcal{N}(v_j)}{w_{jr}}}
% \end{align}
This completes the formation of the weighted line graph from any given network.
% \subsubsection{Weighted Undirected Line Graph}

% Consider two adjacent edges $e_{ij}$ and $e_{jk}$ in the original (undirected) graph $G$. The endpoints of these edges are pair of nodes $(v_i, v_j)$ and $(v_j, v_k)$ respectively. The edge in the $L(G)$ corresponding to $(e_{ij}, e_{jk})$ in $G$ is $e^{L}_{ik} \equiv (e_{ij}, e_{jk}) \equiv ((v_i, v_j), (v_j, v_k))$. We define the edge weight for $e^{L}_{ik}$ as follows:

% {\textcolor{red}{\[ w^L_{ik} = \frac{d_i}{d_i + d_j} \times \frac{w'_{jk}} {\sum_{l=1}^{|V|}{w'_{jl}} - w'_{ij}} \]}}
% % \[ w^L_{ik} = \frac{d_i}{d_i + d_j} \times \frac{w'_{jk}} {\sum_{l=1}^{|V|}{w'_{jl}} - w'_{ij}} \]

% \subsubsection{Weighted Directed Line Graph}

% If the original network $G$ is directed then the line graph $L(G)$ is also directed. Following the same notation used in the case of undirected network, two directed edges $e_{ij}$ and $e_{jk}$ are adjacent if and only if $v_j$ is the destination vertex of $e_{ij}$ and source vertex of $e_{jk}$. We define the edge weight, $w^L_{ik}$, for $e^{L}_{ik}$ as follows:

% {\textcolor{red}{\[ w^L_{ik} = \frac{w'_{jk}} {\sum_{l=1}^{|V|}{w'_{jl}}} \] ** what is j?**}}
% % \[ w^L_{ik} = \frac{w'_{jk}} {\sum_{l=1}^{|V|}{w'_{jl}}} \]

\subsection{Embedding the Line Graph}\label{sec:embedLG}
Here we propose a novel approach to embed the nodes of the line graph. Line graph is a special type of graph which comes with some nice properties. Below is one important observation that we exploit in embedding the line graph.
\begin{lemma}\label{lemma:clique}
\textbf{Each (non-isolated) node in the graph $G$ induces a clique in the corresponding line graph $L(G)$.}
\end{lemma}
\begin{proof}
Let's assume that a (non-isolated) node $v$ in the graph $G$ has $n_v$ edges connected to it. So these $n_v$ edges are neighbors of each other. Hence in the corresponding line graph $L(G)$, each of these edges would be mapped to a node and each of these nodes is connected to all the other $n_v -1$ nodes. Thus there is a clique of size $n_v$ induced in the line graph by node $v$.
\end{proof}
This can be visualized in Fig. \ref{fig:line_graph}, where the node 1 in (a) with degree 3 induces a clique of size 3, including the nodes (1,2), (1,3) and (1,4) into the corresponding line graph in (d).
Lemma \ref{lemma:clique} is interesting because it tells that the nodes of the line graph exhibit some collective property, rather than just pairwise property. To clarify, in the given network, two nodes are pairwise connected by an edge, but in the line graph, a group of nodes form a clique. Pairwise homophily \cite{mcpherson2001birds}, which has been the backbone to many standard embedding algorithms \cite{wang2016structural}, is not sufficient for embedding the line graph.
%This is similar to the idea of going beyond the pairwise similarity, to capture the {\bf gestalt property}, in clustering \cite{watanabe85}.
Hence we propose a new concept `\textbf{collective homophily}' applicable to the line graph. We explain it below.
\begin{figure}%[h]
\centering
\includegraphics[scale=0.25]{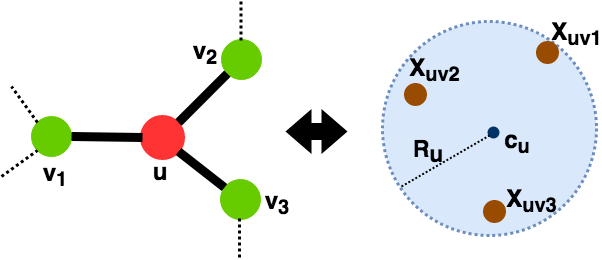}
\caption{Collective Homophily ensures the embeddings of the edges which are connected via a common node in the network, stay within a sphere of \textit{small} radius.}
\label{fig:collecHom}
\end{figure}

\subsubsection{Collective Homophily and Cost Function Formulation}\label{sec:collecHom}
We emphasize that all the nodes, which are part of a clique in a line graph, should be close to each other in the embedding space. One way to enforce collective homophily is to introduce a sphere (of \textit{small} radius $R \in \mathbb{R}$) in the embedding space and ensure that embedding of the nodes (in the line graph) which are part of a clique, remain within the sphere. Hence any two embeddings within a sphere are at a maximum of $2R$ distance apart from each other. The concept is explained in Fig. \ref{fig:collecHom}. Smaller the radius $R$, embeddings of the neighbor edges would be closer to each other and hence the better the enforcement of collective homophily. Note that a sum of pairwise homophily loss in the embedding space %\cite{wang2016structural} 
may lead to some pairs being very close to each other and others may still be quite far. So, we formulate the objective function to embed the (weighted) line graph as follows.

Let us introduce some notation. Bold face letters like $\mathbf{u}$ (or $\mathbf{v}$) denote a node in the line graph $L(G)$, which can also be denoted by $\mathbf{u}_{uv}$ when the correspondence with the edge $(u,v) \in E$ in the original graph $G$ is required. Normal face letters like $u,v$ denote nodes in the given graph. $\mathbf{x}_{\mathbf{v}} \in \mathbb{R}^K$ (equivalently $\mathbf{x}_{uv}$) denotes the embedding of the node $\mathbf{v}_{uv}$ in line graph (or the edge $(u,v) \in E$).

To map the nodes of the line graph to vectors, first we want to preserve different orders of node proximities in the line graph. %As explicitly capturing node proximity is computationally difficult, 
For this, a truncated random walk based sampling strategy $S$ is used to provide a set of nodes $N_S(\mathbf{v})$ as context to a node $\mathbf{v}$ in the network.
%, which helps to capture the proximities. %\cite{perozzi2014deepwalk}. 
Here we employ the random walk proposed by \cite{grover2016node2vec}, which balances between the BFS and DFS search strategy in the graph. As the generated line graph is a weighted one, we consider the weights of the edges while computing the node transition probabilities. Let $X$ denote the matrix with each row as the embedding $\mathbf{x}_\mathbf{v}$ of a node $\mathbf{v}$ of the line graph.
%(or an edge of the original network).
Assuming conditional independence of the nodes, we seek to maximize (w.r.t. $X$) the log likelihood of the context of a node as:
%given its embedding, as:
\begin{align*}
& \sum\limits_{\mathbf{v} \in V_L} \log P(N_S(\mathbf{v})|\mathbf{x}_{\mathbf{v}}) =  \sum\limits_{\mathbf{v} \in V_L} \sum\limits_{\mathbf{v}' \in N_S(\mathbf{v})} \log P(\mathbf{v}'|\mathbf{x}_{\mathbf{v}})
\end{align*}
Each of the above probabilities can be represented using standard softmax function parameterized by the dot product of $\mathbf{x}_{\mathbf{v}'}$ and $\mathbf{x}_{\mathbf{v}}$. As usual, we also approximate the computationally expensive denominator of the softmax function using some negative sampling strategy $\bar{N}(\mathbf{\mathbf{v}})$ for any node $\mathbf{v}$. The above equation, after simple algebraic manipulations, leads to maximizing the following:
\begin{align}\label{eq:opti1}
& 
\sum\limits_{\mathbf{v} \in V_L} 
\sum\limits_{\mathbf{v}' \in N_S(\mathbf{v})} \mathbf{x}_{\mathbf{v}'} \cdot \mathbf{x}_{\mathbf{v}} - |N_S(\mathbf{v})| \; log \Big( \sum\limits_{\bar{\mathbf{v}} \in \bar{N}(\mathbf{v})} exp(\mathbf{x}_{\bar{\mathbf{v}}} \cdot \mathbf{x}_{\mathbf{v}})\Big)
\end{align}

Next, we implement the concept of collective homophily as proposed above. Each node $u \in V$ (in the original network) induces a clique in the line graph (Lemma \ref{lemma:clique}). An edge $(u,v) \in E$ corresponds to the node $\mathbf{v}_{uv} \in V_L$ in the line graph. So we want all the nodes of the form $\mathbf{v}_{uv} \in V_L$ belong to a sphere centered at $\mathbf{c}_u \in \mathbb{R}^K$ and of radius $R_u$, where $v \in \mathcal{N}(u)$ (neighbors of $u$). As collective homophily suggests that embeddings of these nodes must be close to each other, we minimize the sum of all such radii. This with Eq. \ref{eq:opti1} gives the final cost function of line2vec as follows. 
\begin{equation}\label{eq:optiCH}
\begin{aligned}
& \underset{X,R,C}{\text{min}}
& & \sum\limits_{\mathbf{v} \in V_L} \Big[ |N_S(\mathbf{v})| \; log \Big( \sum\limits_{\bar{\mathbf{v}} \in \bar{N}(\mathbf{v})} exp(\mathbf{x}_{\bar{\mathbf{v}}} \cdot \mathbf{x}_{\mathbf{v}})\Big) \\
&&& - \sum\limits_{\mathbf{v}' \in N_S(\mathbf{v})} \mathbf{x}_{\mathbf{v}'} \cdot \mathbf{x}_{\mathbf{v}} \Big] + \alpha \sum\limits_{u \in V} R_u^2\\
& \text{such that,}
& & || \mathbf{x}_{uv} - \mathbf{c}_{u} ||_2^2 \leq R_u^2, \; \forall v \in \mathcal{N}(u), \; \forall u \in V \\
&&& R_u \geq 0, \; \forall u \in V
\end{aligned}
\end{equation}

Here, $\alpha$ is a positive weight factor. The constraint $|| \mathbf{x}_{uv} - \mathbf{c}_{u} ||_2^2 \leq R_u^2$ ensures that nodes of the form $\mathbf{x}_{uv}$ belong to the sphere of radius $R_u$ and centered at $\mathbf{c}_{u}$. We use $R$ and $C$ to denote set of all such radii and centers respectively.

\subsubsection{Solving the Optimization}\label{sec:solOpti}
Equation \ref{eq:optiCH} is a non-convex constrained optimization problem. We use penalty functions \cite{bryan2005penalty} technique to convert this to an unconstrained optimization problem as follows:
\begin{equation}\label{eq:optiPen}
\begin{aligned}
& \underset{X,R,C}{\text{min}}
\;\;\; \sum\limits_{\mathbf{v} \in V_L} \Big[ |N_S(\mathbf{v})| \; log \Big( \sum\limits_{\bar{\mathbf{v}} \in \bar{N}(\mathbf{v})} exp(\mathbf{x}_{\bar{\mathbf{v}}} \cdot \mathbf{x}_{\mathbf{v}})\Big) \\
& - \sum\limits_{\mathbf{v}' \in N_S(\mathbf{v})} \mathbf{x}_{\mathbf{v}'} \cdot \mathbf{x}_{\mathbf{v}} \Big] + \alpha \sum\limits_{u \in V} R_u^2 \\
& + \lambda \sum\limits_{u \in V} \sum\limits_{v \in \mathcal{N}(u)} g(|| \mathbf{x}_{uv} - \mathbf{c}_{u} ||_2^2 - R_u^2)
+ \sum\limits_{u \in V} \gamma_u g(-R_u)
\end{aligned}
\end{equation}
Here the function $g : \mathbb{R} \rightarrow \mathbb{R}$ is defined as $g(t) = max(t,0)$. So it imposes a penalty to the cost function in Eq. \ref{eq:optiPen} when the argument inside $g$ is positive, i.e., when there is a violation of the constraints in Eq. \ref{eq:optiCH}. We use a linear penalty $g(t)$ as the gradient does not vanish even when $t \to 0_+$. To solve the unconstrained optimization in Eq. \ref{eq:optiPen}, we use stochastic gradient descent, computing gradients w.r.t. each of $X$, $R$ and $C$. We take subgradient when $t=0$ for $g(t)$. All the  penalty parameters $\lambda$ and $\gamma_u$'s corresponding to penalty functions are positive. When there is any violation of a constraint (or sum of constraints), the corresponding penalty parameter is increased to impose more penalty.
We give more importance to the type of constraints $R_u \geq 0$, as violation of them may change the intuition of the solution. So we use different penalty parameters for each of them, so imposing a different penalty to each of such constraints is possible.
One can show that under appropriate assumptions, any convergent subsequence of solutions to the unconstrained penalized problems must converge to a solution of the original constrained problem \cite{bryan2005penalty}.
Very small values of the penalty parameters might lead to the violation of constraints, and very large values would make the gradient descent algorithm oscillate. So, we start with smaller values of $\lambda$ and $\gamma_u$'s and keep increasing them until all the constraints are satisfied or the gradients become too large making abrupt function changes. Note that, theoretically some of the constraints in Eq. \ref{eq:optiCH} may still be violated, but experimentally we found them satisfied up to a large extent (Section \ref{sec:expEva}). In the final solution, $\mathbf{x_v}$ gives the vector representation of node $\mathbf{v}$ of the line graph, which is essentially the embedding of the corresponding edge in the original network.

\subsection{Key Observations and Analysis}% on the Transformed Line graph}
%Pseudo-code of line2vec is presented in the supplementary material.
Both the edge embedding properties mentioned in Section \ref{sec:prob} are preserved in the construction and embedding of the weighted line graph. Particularly, if two edges have a common incident node in the original network, the corresponding two nodes in the transformed line graph would be neighbors. Also two edges having similar neighborhood in the original network lead to two nodes having similar neighborhood in the transformed line graph. The random walk and collective homophily preserve both pairwise and collective node proximity of the line graph in the embedding space. Thus different orders of edge proximities of the original network is captured well in the edge embeddings. Also the construction of edge weights in line graph (Sec. \ref{sec:weightLG}) ensures that underlying importance of edges of the original network is preserved in the transformed line graph, and hence in the embeddings through truncated random walk. 

\textbf{Time Complexity}: Edge embedding is computationally difficult than node embedding, as the number of edges in a real life network is more than the number of nodes. %First let us calculate the number of edges in line graph $L(G)$, for the given graph $G=(V,E)$. %
From Lemma \ref{lemma:clique}, each node $u$ in the original network induces a clique of size $d_u$ (degree of $u$ in $G$). Hence total number of edges in the line graph is: $m_L$ = $\sum\limits_{u \in V} {d_u \choose 2}$ = $\sum\limits_{u \in V} \frac{d_u (d_u - 1)}{2}$ $\leq$ $|V| d^2$, where $d$ is the maximum degree of a node in the given network. So, the construction of line graph would take $O(|V| d^2)$ time.
%As mentioned in Section \ref{sec:collecHom}, we use a random walk to generate the context of each node in the line graph. 
Next, we use alias table for fast computation of the corpus of node sequences in $O(m_L \; log(m_L))$ = $O(|V| d^2 \; log(|V| d))$ by the random walks, assuming the number of random walks on the line graph, maximum length of a random walk, context window size and the number of negative samples for each node to be constant, as they are the hyper parameters of skip-gram model.
Then, the first term (under the sum over the nodes in $V_L$) of Eq. \ref{eq:optiPen} can be computed in $O(|V_L|)=O(|E|)$ time. Next, the term weighted by $\alpha$ can be computed in $O(|V|)$ time. Then, for the term weighted by $\lambda$, we need to visit each node in $V$ and for each such node, its neighbors in the original graph, which can be computed in a total of $O(|E|)$ time. The last term of Eq. \ref{eq:optiPen} can be computed in $O(|V|)$ time. As we use penalty methods to solve it, the runtime of solving Eq. \ref{eq:optiPen} is $O(|E|+|V|)$.
Hence the total runtime complexity of line2vec is 
%$O(|V| d^2 \; + \; |V| d^2 \; log(|V| d) \; + \; |E| + |V|)$ = 
$O(|V| d^2 \; log(|V| d))$. So in the worst case, (for e.g., a nearly complete graph), run time complexity is $O(|V|^3 log |V|)$. But for most of the real life social networks, the maximum degree can be considered as a constant (i.e., does not grow with the number of nodes). Hence for them, the run time complexity is $O(|V| log |V|)$.

\begin{figure*}[h!]
  \centering
  \begin{subfigure}[b]{0.16\linewidth}
    \includegraphics[width=\linewidth]{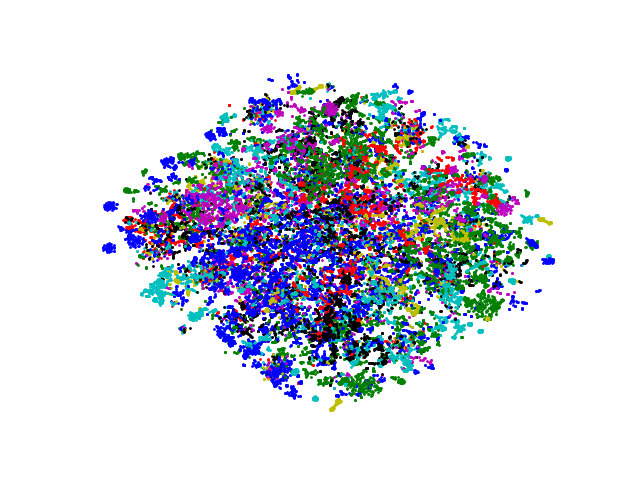}
    \caption{DeepWalk}
  \end{subfigure}
  \begin{subfigure}[b]{0.16\linewidth}
    \includegraphics[width=\linewidth]{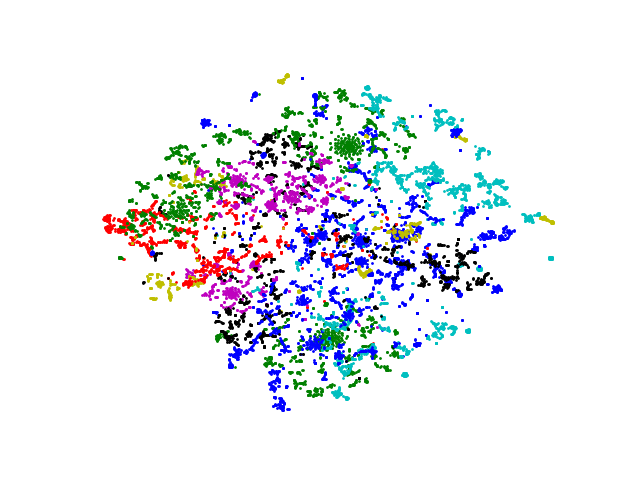}
    \caption{node2vec}
  \end{subfigure}
  \begin{subfigure}[b]{0.16\linewidth}
    \includegraphics[width=\linewidth]{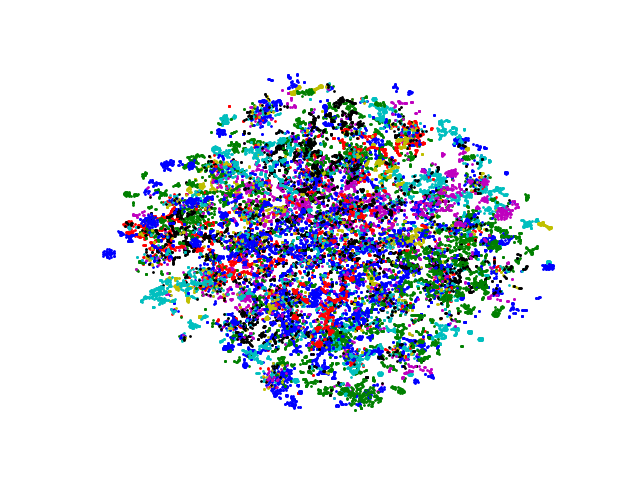}
    \caption{SDNE}
  \end{subfigure}
  \begin{subfigure}[b]{0.16\linewidth}
    \includegraphics[width=\linewidth]{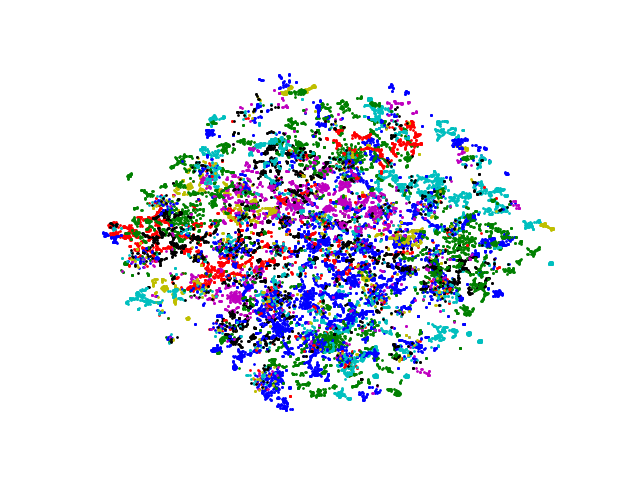}
    \caption{struc2vec}
  \end{subfigure}
  \begin{subfigure}[b]{0.16\linewidth}
    \includegraphics[width=\linewidth]{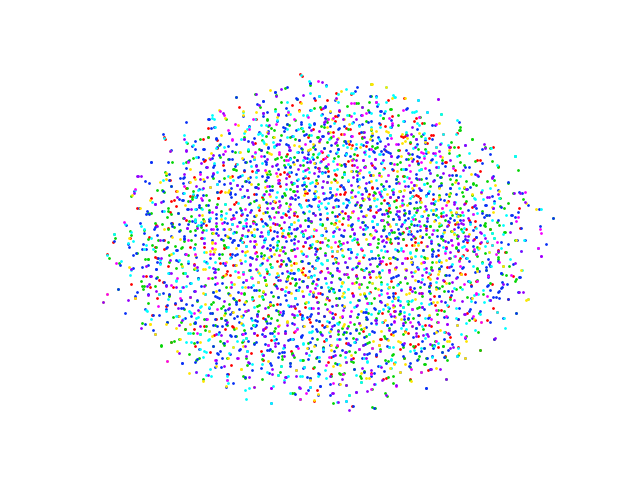}
    \caption{GraphSAGE}
  \end{subfigure}
  \begin{subfigure}[b]{0.16\linewidth}
    \includegraphics[width=\linewidth]{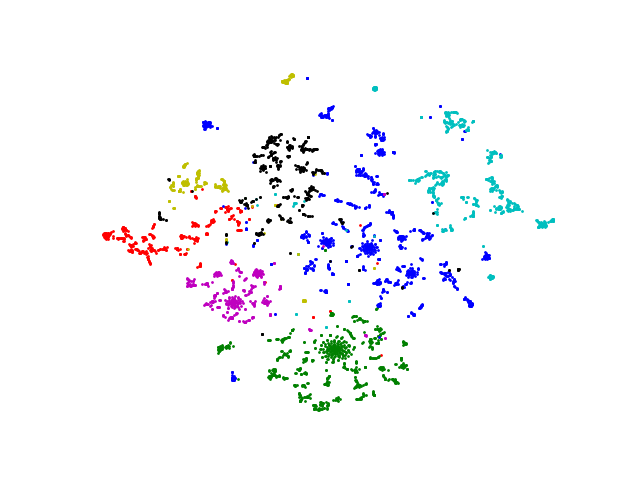}
    \caption{line2vec}
  \end{subfigure}
  \caption{Edge visualization on Cora dataset. Different colors represent different edge communities.}
  \label{fig:visCora}
\end{figure*}

\section{Experimental Evaluation}\label{sec:expEva}
We conduct detailed experiments on three downstream edge centric network mining tasks and thoroughly analyze the proposed optimization framework of line2vec.

\subsection{Design of Baseline Algorithms}\label{sec:baselines}
%As stated before, 
Unsupervised direct edge embedding for information network itself is a novel problem. Existing approaches only aggregate the embeddings of the two end nodes to find the embedding of an edge. So as baselines, we only consider the publicly available implementation of a  set of popular unsupervised node embedding algorithms which can work only using the link structure of the graph: DeepWalk, node2vec, SDNE, struc2vec and GraphSage (official unsupervised implementation for the un-attributed networks).
We have considered different types of node aggregation methods such as taking the average, Hadamard product, vector norms of two end node embeddings \cite{grover2016node2vec} to generate the edge embeddings for the baseline algorithms. It turns out that average aggregation method performs the best among them. So we report the performance of the baseline methods with average node aggregation, where embedding of an edge $(u,v)$ is computed by taking the average of the node embeddings of $u$ and $v$. %, computed by the baseline.

\subsection{Datasets Used and Setting Hyper-parameters}\label{sec:datasets}
We used five real world publicly available datasets for the experiments. A summary of the datasets is given in Table \ref{tab:data}. For Zachary's karate club and Dolphin social network (\url{http://www-personal.umich.edu/~mejn/netdata/}),
%\footnote{\label{foot:ds1}\url{http://www-personal.umich.edu/~mejn/netdata/}}, 
there are no ground truth community labels given for the nodes. So we use the modularity based community detection algorithm, and label the nodes based on the communities they belong to. For Cora, Pubmed (\url{https://linqs.soe.ucsc.edu/data})
%\footnote{\label{foot:ds2}\url{https://linqs.soe.ucsc.edu/data}} 
and MSA \cite{sinha2015overview}, the ground truth node communities are available.
%Once we have the community labels for the nodes, 
The ground truth edge labels are derived as follows. If an edge connects two nodes of the same community (intra community edge), the label of that edge is the common community label. If an edge connects nodes of different communities (inter community edge), then that edge is not considered for calculating the accuracy of downstream tasks. 
%The number of ground truth community labels of the edges, as derived above, is mentioned in Table \ref{tab:data} and an edge can belong to only one community. 
Note that, all the edges (both intra and inter community) are considered for learning the edge embeddings.
We also provide the size of the generated weighted line graphs in Table \ref{tab:data}. 
Note that, line graphs are still extremely sparse in nature, which enables the application of efficient data structures and computation on sparse graphs here. %For e.g., average degree of a node in the line graph of MSA is only ~30, and that in Pubmed is only ~15. 
%Hence efficient data structures and computation available for sparse graphs can easily be used for line graphs also.
\begin{table}
    \caption{Summary of the datasets used.}% \#Labels denotes the number of ground truth communities (for nodes or edges) in the network, which is derived for the first two networks.}
	\centering
 	%\footnotesize
    \resizebox{\columnwidth}{!}{%
	\begin{tabular}{c c c c | c c}
	\toprule
	\sffamily{Dataset} & \#Nodes & \#Edges & \#Edge-Labels & \#Nodes in $L(G)$ & \#Edges in $L(G)$ \\
%    \sffamily{} & & & & Words & Distribution & links \\
    \hline
	\midrule
    \sffamily{Zachary's Karate club} & 34 & 78  & 3 & 78 & 528\\
    \sffamily{Dolphin social network} & 62 & 159 & 4 & 159 & 923\\
    \sffamily{Cora} & 2708 & 5278 & 7 & 5278 & 52301 \\
    \sffamily{Pubmed} & 19717 & 44327 & 3 & 44327 & 699385\\
    \sffamily{MSA} & 30101 & 204926 & 3 & 204926 & 6149555\\ 
\bottomrule
	\end{tabular}
    }
	%\vspace{-5mm}
	\label{tab:data}
	\end{table} %sam modi 

%\subsection{Hyper-parameter Setup}
We set the parameter $\alpha$ in Eq. \ref{eq:optiCH} to be 0.1 in the experiments. At that value, the two components in the cost function in Eq. \ref{eq:optiCH} contribute roughly the same to the total cost in the first iteration of line2vec. The dimension ($K$) of the embedding space is set as 8 for Karate club and Dolphin social network as they are small in size, and it is set as 128 for the other three larger datasets (for all the algorithms). For the faster convergence of SGD, we set the initial learning rate higher and decrease it over the iterations.
%
%As mentioned before, penalty functions do not guarantee the satisfiability of the constraints in practice. But 
We vary the penalty parameters in Eq. \ref{eq:optiPen} over the iterations as discussed in Section \ref{sec:solOpti} to ensure that the constraints are satisfied at large.

\subsection{Penalty Errors of line2vec Optimization}
\begin{figure}%[h]
\centering
\begin{subfigure}{0.49\columnwidth}
\includegraphics[width=0.95\linewidth]{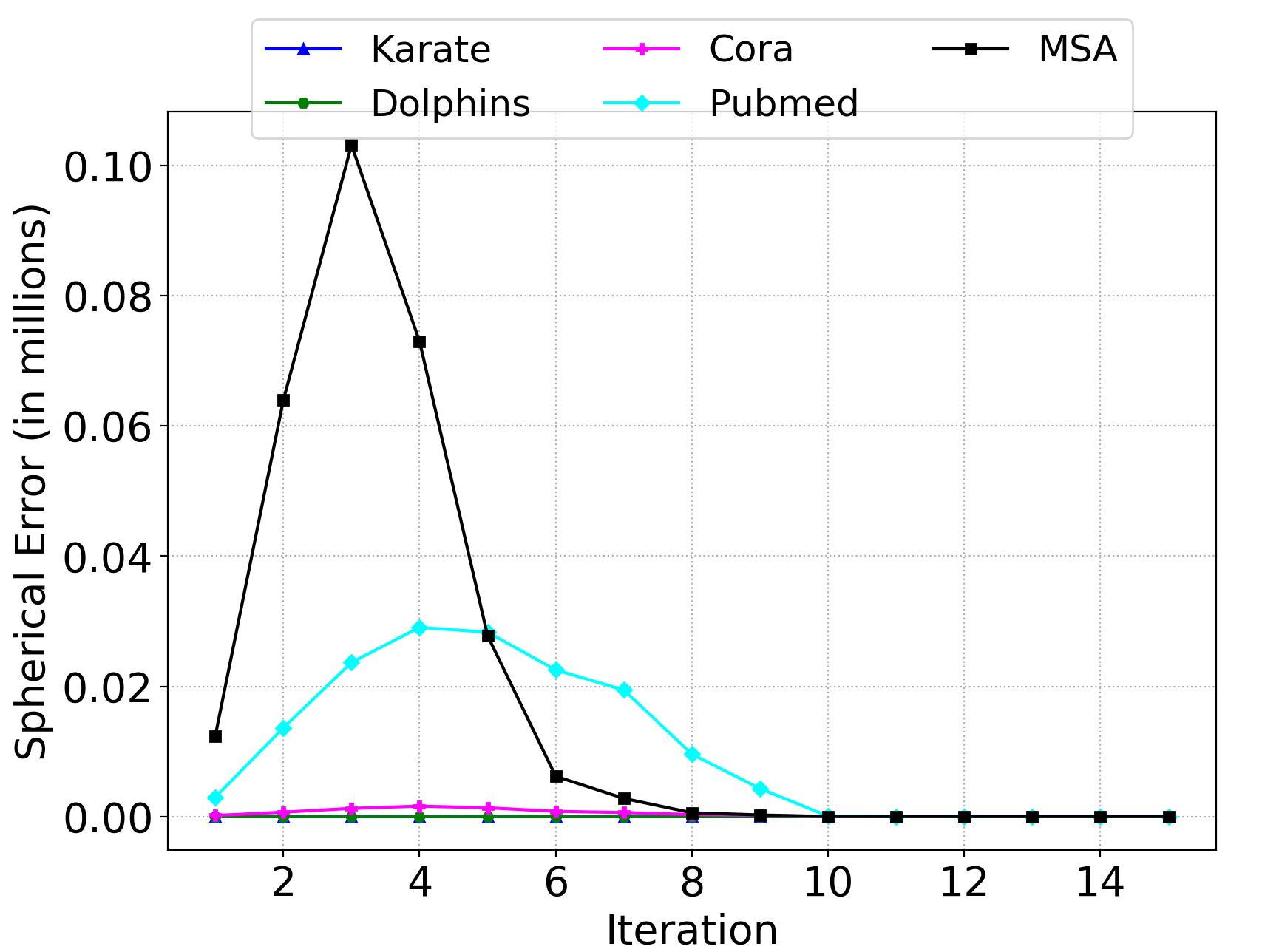}
\captionsetup{font=scriptsize}
\caption{Spherical Error}
\end{subfigure}
\begin{subfigure}{0.49\columnwidth}
\includegraphics[width=0.95\linewidth]{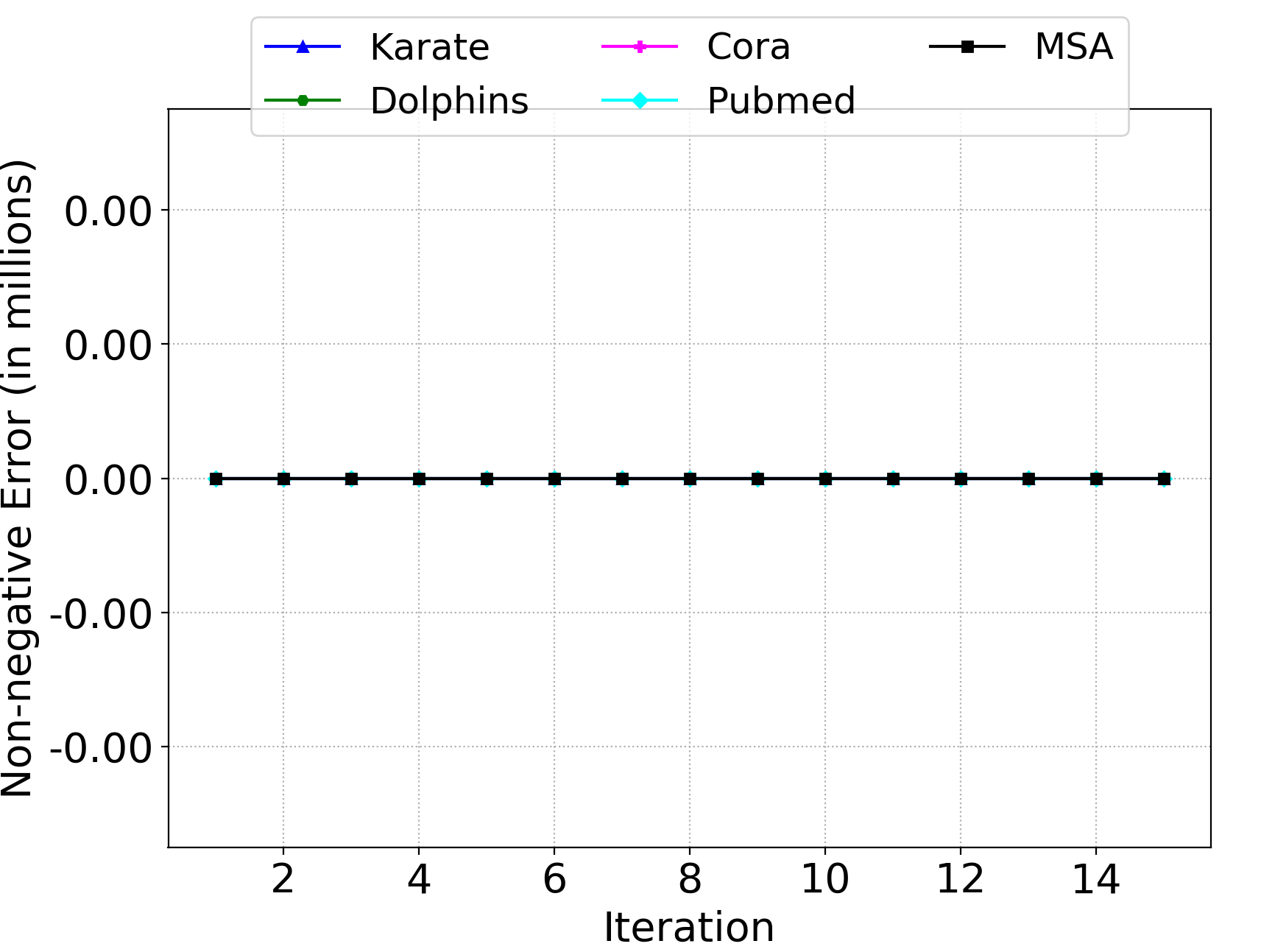}
\captionsetup{font=scriptsize}
\caption{Non-negative Error}
\end{subfigure}

\caption{Both spherical error $\sum\limits_{u \in V} \sum\limits_{v \in \mathcal{N}(u)} g(|| x_{uv} - \mathbf{c}_{u} ||_2^2 - R_u^2)$ and non-negative error  $\sum\limits_{u \in V} g(-R_u)$ in the penalty function based optimization of line2vec converge to zero very fast on all the datasets.}
\label{fig:error}
\end{figure}

We have shown the values of two different penalty errors (or constraint violation error of the penalty method based optimization) over the iterations of line2vec in Figure \ref{fig:error}. 
For all the datasets, total spherical error $\sum\limits_{u \in V} \sum\limits_{v \in \mathcal{N}(u)} g(|| x_{uv} - \mathbf{c}_{u} ||_2^2 - R_u^2)$ converges to a small value very close to zero and negative error $\sum\limits_{u \in V} g(-R_u)$ remains to be zero.
This means, almost all the constraints of line2vec formulation are satisfied in the final solution.

\subsection{Downstream Edge Mining Tasks}\label{sec:downstream}
%\subsection{Edge Visualization}\label{sec:vis}
% We consider three important downstream edge mining tasks to show the usefulness of line2vec to generate edge embeddings.
%
\textbf{Edge visualization}: It is important to understand if the edge embeddings are able to separate the communities visually. We use the embedding of the edges as input in $\mathbb{R}^K$, and use t-SNE \cite{maaten2008visualizing} to plot the edge embedding in a 2 dimensional space.
Fig. \ref{fig:visCora} shows the edge visualizations by line2vec, along with the baselines algorithms on Cora datasets. Note that, line2vec is able to visually separate the communities well compared to all the other baselines. The same trend was observed even in Fig. \ref{fig:motiv} for the small synthetic network. Line2vec, being a direct approach for edge embedding via collective homophily, outperforms all the baselines which aggregate node embeddings to generate the embeddings for the edges.
%{\textcolor{red}{Check numbering of Figures}}

\begin{figure*}[h!]
  \begin{subfigure}[b]{0.32\linewidth}
  \centering
    \includegraphics[width=\linewidth]{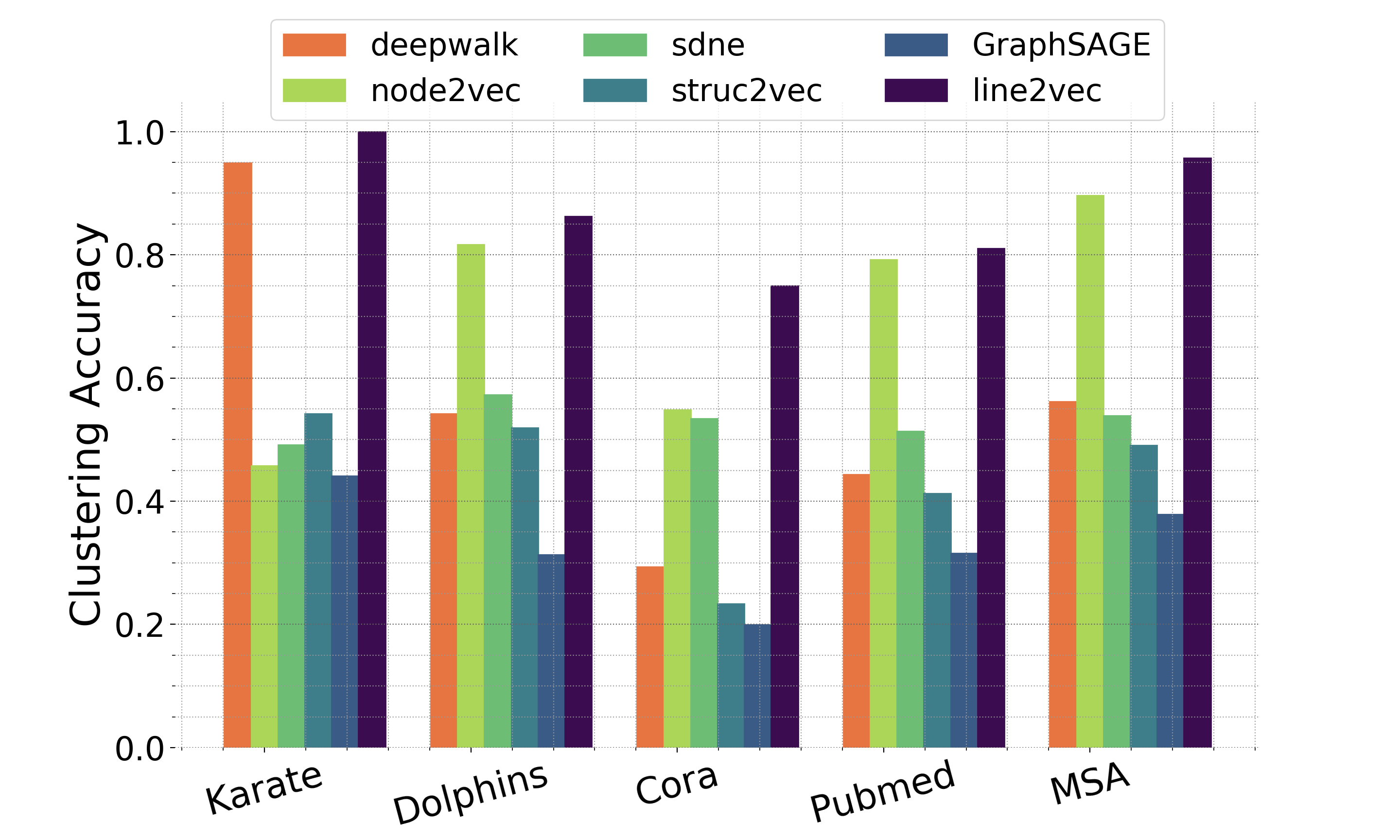}
    \caption{Edge Clustering}
    \label{fig:clustering}
  \end{subfigure}
  \begin{subfigure}[b]{0.32\linewidth}
  \centering
    \includegraphics[width=\linewidth]{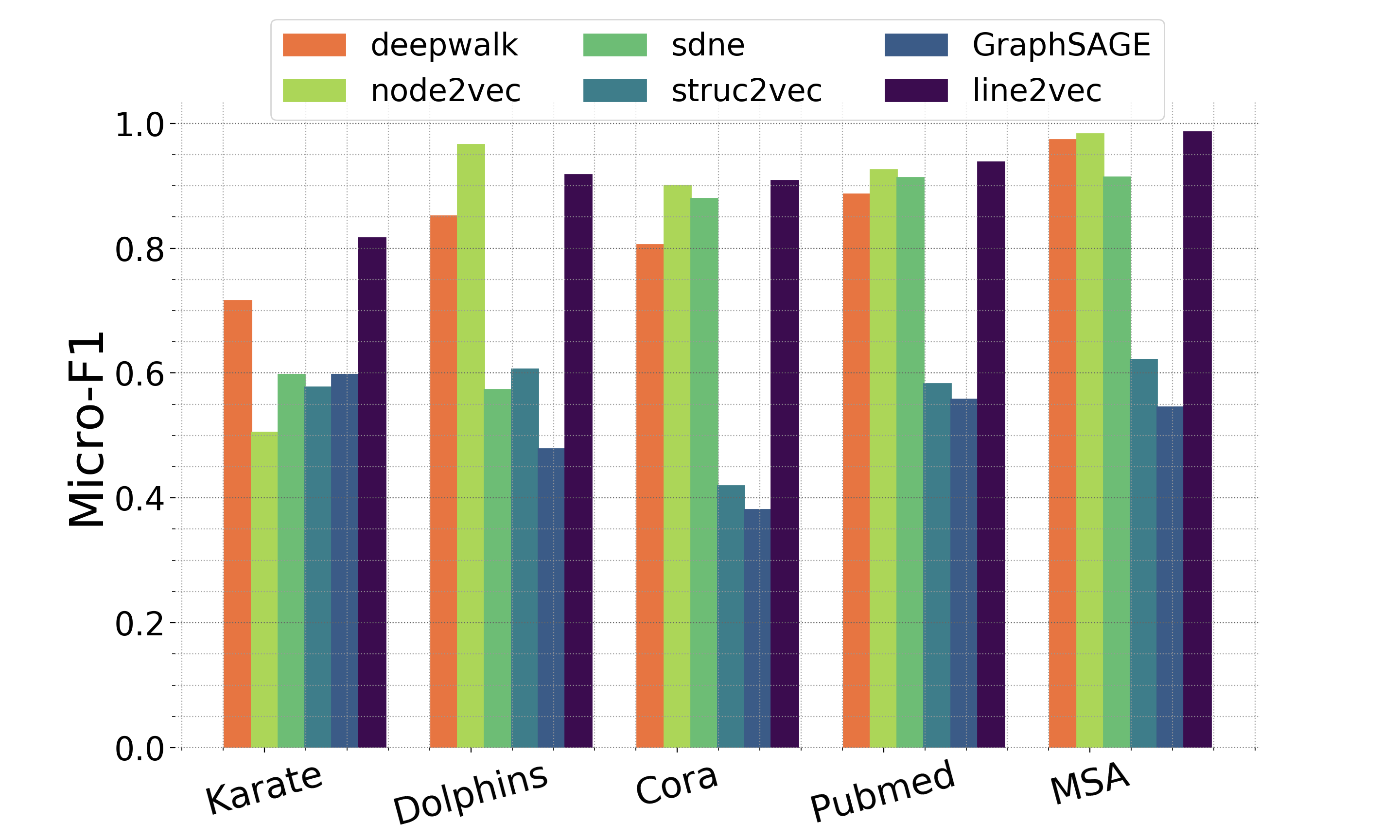}
    \caption{Edge Classification}
    \label{fig:Micclassification}
  \end{subfigure}
  \begin{subfigure}[b]{0.32\linewidth}
  \centering
    \includegraphics[width=\linewidth]{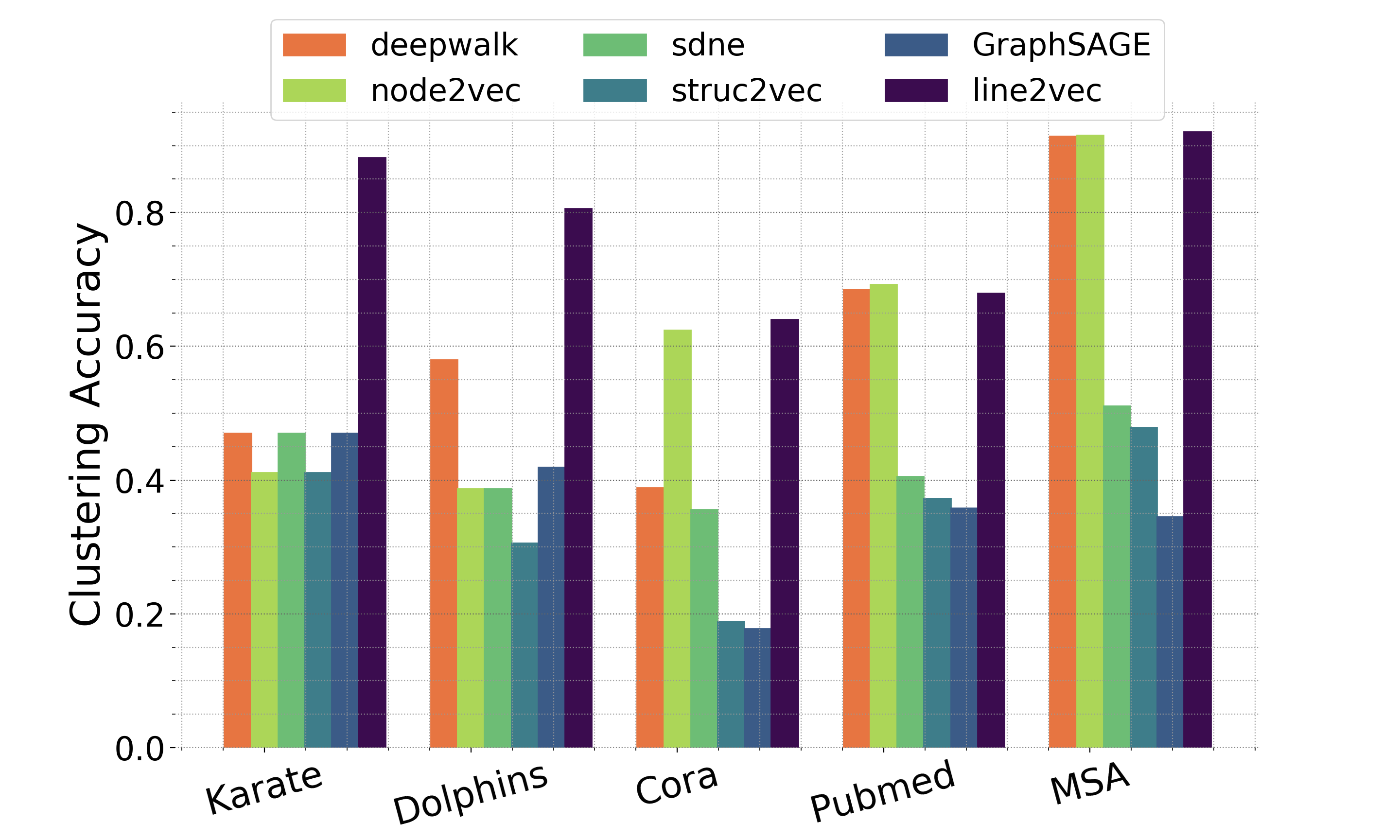}
    \caption{Node Clustering}
    \label{fig:nodeClustering}
  \end{subfigure}
  \caption{Performance Comparisons: (a) Micro F1 Score of Edge Classification. (b) Edge Clustering with KMeans++. (c) Node Clustering with KMeans++. Here we use $\mathbf{c_u}$ as the embedding of the node $u$ in the given network.}
  \label{fig:downstream}
\end{figure*}

% \begin{figure}%[h]
% \centering
% \includegraphics[scale=0.22]{images/micro.png}
% \caption{Micro F1 Score of Edge Classification}
% \label{fig:Micclassification}
% \end{figure}

%\subsection{Edge Clustering}\label{sec:clus}
%Node clustering or node community detection are popular network mining tasks. In this paper, for the first time in literature, 
%We address edge clustering in information networks via edge embedding for the first time in literature. 
\textbf{Edge Clustering}: Like node clustering, edge clustering is also important to understand the flow of information within and between the communities. 
For clustering the embeddings of the edges, we apply KMeans++ algorithm.
To evaluate the quality of clustering, we use unsupervised clustering accuracy \cite{xie2016unsupervised} which uses different permutations of the labels and chooses the assignment which gives best possible accuracy. 
Figure \ref{fig:clustering} shows that line2vec outperforms all the baselines for edge clustering on all the datasets. DeepWalk and node2vec also perform well among the baselines.

%\subsection{Multi-class Edge Classification}
\textbf{Multi-class Edge Classification}: We use only 10\% edges with ground truth label (as generated in Section \ref{sec:datasets}) as the training set, because getting labels is expensive in networks. A logistic regression classifier is trained on the edge embeddings generated by different algorithms. The performance on the test set is reported using Micro F1 score. Figure \ref{fig:Micclassification} shows that line2vec is better or highly competitive with the state-of-the-art embedding algorithms. node2vec and DeepWalk follows line2vec closely. On the Dolphin dataset, node2vec outperforms line2vec marginally. Performance of line2vec for edge classification again shows the superiority of a direct edge embedding scheme over the node aggregation approaches.
%Note that the performance of baselines depends heavily on the aggregation method and we reported the best accuracy numbers with average aggregation. Whereas, line2vec is focused on generating edge embeddings directly and does not need any aggregation.

% \begin{figure}%[h]
% \centering
% \includegraphics[scale=0.2]{images/clus_acc.png}
% \caption{Edge Clustering with KMeans++}
% \label{fig:clustering}
% \end{figure}

\subsection{Ablation Study of line2vec}\label{sec:incre_l2v}
\begin{figure}[h!]
  \centering
  \begin{subfigure}[b]{0.23\linewidth}
    \includegraphics[width=\linewidth]{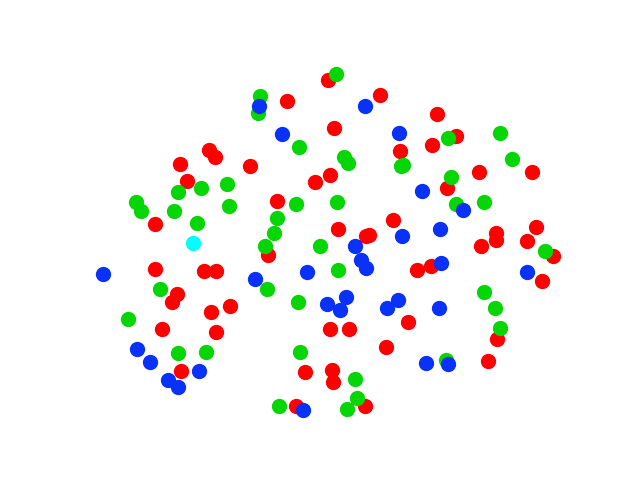}
    \caption{N2V}
    \label{subfig:visDoln2v}
  \end{subfigure}
  \begin{subfigure}[b]{0.23\linewidth}
    \includegraphics[width=\linewidth]{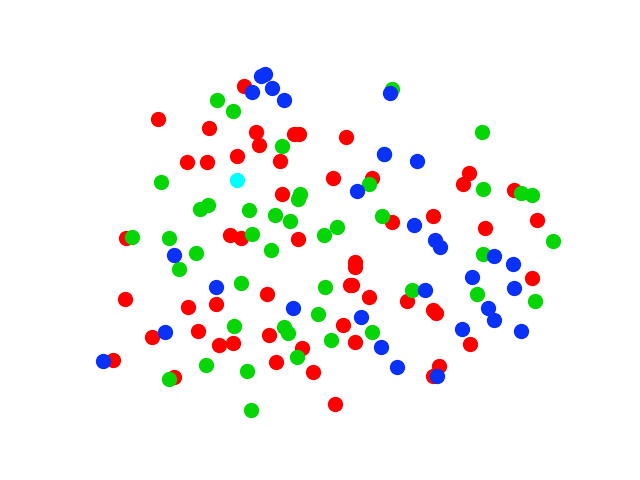}
    \caption{N2V+LG}
    \label{subfig:visDoln2v+lg}
  \end{subfigure}
  \begin{subfigure}[b]{0.23\linewidth}
    \includegraphics[width=\linewidth]{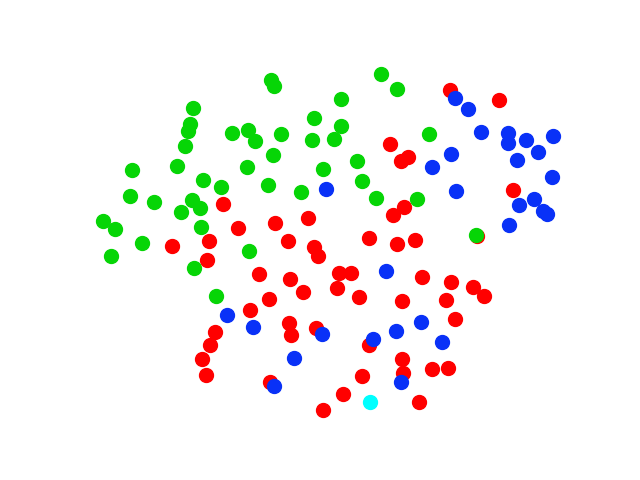}
    \caption{N2V+WLG}
    \label{subfig:visDoln2v+wlg}
  \end{subfigure}
  \begin{subfigure}[b]{0.23\linewidth}
    \includegraphics[width=\linewidth]{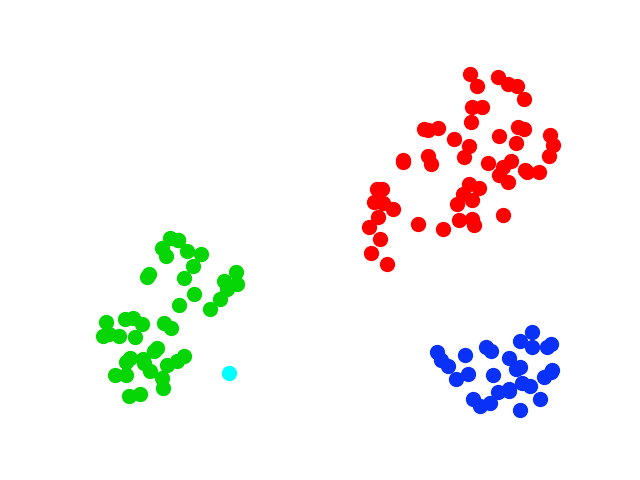}
    \caption{L2V}
    \label{subfig:visDoll2v}
  \end{subfigure}
  \caption{Edge visualization on Dolphin Dataset by t-SNE: In the following sub-figures, edge Embeddings are obtained \textbf{(a)} by using node2vec on the input graph and then taking average of end node embeddings for each edge, \textbf{(b)} by using node2vec on an unweighted (conventional) line graph, \textbf{(c)} by using node2vec on our proposed weighted line graph, \textbf{(d)} by line2vec. Clearly, there is an \textbf{incremental improvement} of the quality because of using weighted line graph and then collective homophily as reflected in (c) and (d) respectively.}
  %line2vec performs the best as it uses skip-gram based objective on weighted line graph, along with collective homophily on top of that.}
  \label{fig:incre_l2v}
\end{figure}

The idea of line2vec is to embed the line graph for generating the edge embeddings of a given network. There are two main novel components in line2vec: first, the construction of weighted line graph; and second, more importantly, proposing the concept of collective homophily on the weighted line graph. In this subsection, we show the incremental benefit of each component through a small experiment of edge visualization on the Dolphin dataset, as shown in Fig. \ref{fig:incre_l2v}. We use node2vec (N2V) as the starting point because the skip-gram objective component of line2vec (L2V) is similar to node2vec. Though, visually there is not much difference between Sub-figures \ref{subfig:visDoln2v} and \ref{subfig:visDoln2v+lg}, but there is some improvement when we apply node2vec on the weighted line graph (without using collective homophily) in Sub-fig. \ref{subfig:visDoln2v+wlg}. Finally, superiority of line2vec because of using collective homophily on the weighted line graph is clear from Sub-fig. \ref{subfig:visDoln2v+wlg}.
%where all the communities are separated distinctly. 
Thus, both the novel components of line2vev have their incremental benefits for the overall algorithm.

% \begin{figure}%[h]
% \centering
% \includegraphics[scale=0.22]{images/node_clus_acc.png}
% \caption{Node Clustering with KMeans++: Here we use $\mathbf{c_u}$ as the embedding of the node $u$ in the given network.}
% \label{fig:nodeClustering}
% \end{figure}

\subsection{Parameter Sensitivity of line2vec}\label{sec:para_sensitivity}
\begin{figure}[h!]
    \small
  \centering
  \begin{subfigure}[b]{0.45\linewidth}
    \includegraphics[width=0.9\linewidth]{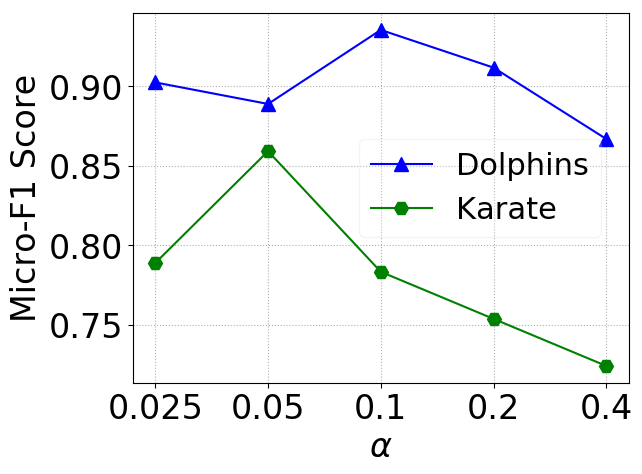}
    \caption{Edge Classification}
  \end{subfigure}
%   \begin{subfigure}[b]{0.32\linewidth}
%     \includegraphics[width=\linewidth]{images/alpha_macro.png}
%     \caption{}
%   \end{subfigure}
  \begin{subfigure}[b]{0.45\linewidth}
    \includegraphics[width=0.9\linewidth]{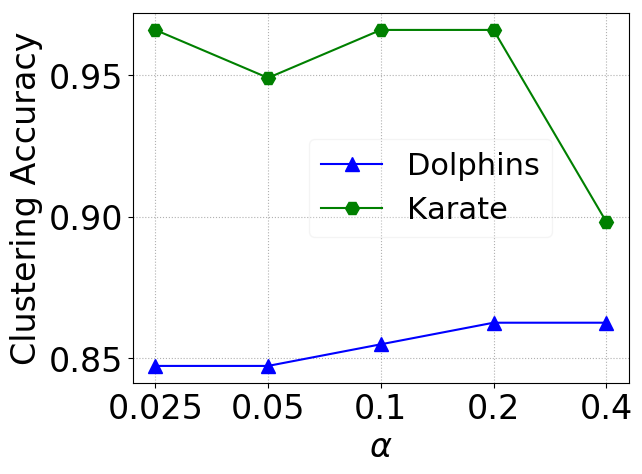}
    \caption{Edge Clustering}
  \end{subfigure}
  \caption{Sensitivity of line2vec with respect to the hyper-parameter $\alpha$ (in Eq. 3 of the main paper) on Karate and Dolphin datasets: We have shown the variation of performance for edge classification (Micro F1 score) and edge clustering (unsupervised accuracy).}
  \label{fig:parameter_sen}
\end{figure}

Figure \ref{fig:parameter_sen} shows the sensitivity of line2vec with respect to the hyper-parameter $\alpha$ (in Eq. 3 of the main paper) on Karate and Dolphin datasets. We have shown the variation of performance for node classification (both micro and macro F1 scores) and node clustering (unsupervised accuracy). From the figure, one can observe that optimal performance in most of the cases is obtained when the value of $\alpha$ is from 0.05 to 0.1. Around these values, the loss from both the components of line2vec in Eq. 3 are close to each other. For our other experiments, we fix $\alpha$=0.1 for all the datasets.

\subsection{Interpretation of $\mathbf{c_u}$ as Node Embedding}
line2vec is dedicated for direct edge embedding in information networks. Lemma \ref{lemma:clique} suggests that each node in the given network $G$ induces a clique in the line graph $L(G)$. Based on the concept of collective homophily, corresponding to a node $u$ in $G$, the clique in the line graph is enclosed by a sphere centered at $\mathbf{c}_u \in \mathbb{R}^K$
%and with radius $R_u$ 
(Eq. \ref{eq:optiCH}). 
Intuitively, the center acts as a point which is close to the embeddings of all the nodes in the clique induced by $u$ (or equivalently, all the edges incident on $u$ in $G$). Hence the role of this center in the embedding space is similar to the role of the node to its adjacent edges in the graph. This motivates us to consider $\mathbf{c}_u$ as the node embedding of $u \in V$ in $G$. If $(u,v) \in E$, then the edge embedding of $(u,v)$ should be close to both $\mathbf{c_u}$ and $\mathbf{c_v}$, which in turn pulls $\mathbf{c_u}$ and $\mathbf{c_v}$ close to each other. Thus, node proximities are also captured in $\mathbf{c_u}$.%, $\forall u$.

We use clustering of the nodes (a.k.a. community detection) of the given network to validate the quality of node embedding obtained from the centers of the line2vec optimization. We use k-means++ clustering, as before, on the set of points $\mathbf{c}_u$, $\forall u \in V$ and validate the clustering quality by using unsupervised accuracy \cite{bandyopadhyay2019outlier}. 
%Among the datasets we used, Cora, Pubmed and MSA have the ground truth labels of the nodes, which we considered as their true communities. For Karate club and Dolphin network, similar to the Section \ref{sec:datasets}, we use the communities obtained from modularity based community detection algorithm as the ground truth. We again use unsupervised clustering accuracy \cite{xie2016unsupervised} to judge the quality of clustering.
Figure \ref{fig:nodeClustering} shows that line2vec, though designed specifically for edge embedding, performs really good for a node based mining task. Specifically, for Karate and Dolphins networks, the gain is significantly more than best of the baselines. 
%For other three datasets, line2vec performs almost as good as the best baseline. 
This result is interesting as we aimed to find edge embeddings, but also generate a set of efficient node embeddings, which are not just the aggregation of the incident edges.

\subsection{Connection of Node Centrality with $R_u$}\label{sec:inter}
This subsection analyzes the interpretation of the radius $R_u$ of the sphere enclosing the clique induced by node $u \in V$ in the embedding space. When a node $u$ has less number of incident edges, and the neighbors are very close to each other in the embedding space (for e.g., they are all from the same sub-community), a small radius $R_u$ should be enough to enclose all the edges incident on $u$. But when the neighbors of the node $u$ are diverse in nature, the corresponding edges would also be different in terms of strength and semantics.
For example, an influential researcher may be directly connected to many other researchers in a research network, but only few of them can be direct collaborators. Hence, a larger sphere is needed to enclose the clique in the line graph induced by such a node. This intuition connects radius $R_u$ of a sphere in the embedding space of line graph to the centrality \cite{skibski2016attachment} of the node $u$ in the given network. A node which is loosely connected (i.e., less number or very similar neighbors) in the network is less central, and a node which is strongly connected (many or diverse set of neighbors) is considered as highly central.
%Following experiments validate if $R_u$ is correlated with the centrality of the node $u \in V$. 
%

\begin{figure}[h!]
    \small
  \centering
  \begin{subfigure}[b]{0.32\linewidth}
    \includegraphics[width=\linewidth]{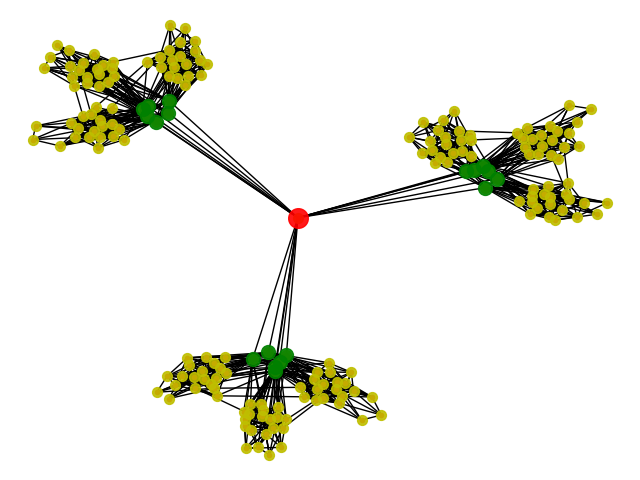}
    \caption{}
  \end{subfigure}
  \begin{subfigure}[b]{0.32\linewidth}
    \includegraphics[width=\linewidth]{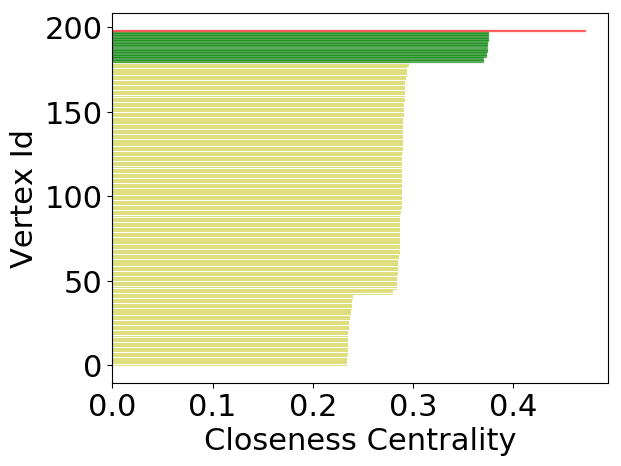}
    \caption{}
  \end{subfigure}
  \begin{subfigure}[b]{0.32\linewidth}
    \includegraphics[width=\linewidth]{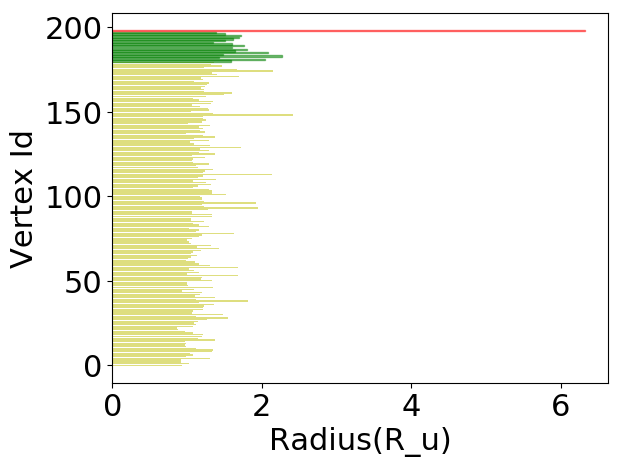}
    \caption{}
  \end{subfigure}
  \caption{Relationship between radius $R_u$ associated with each node and closeness centrality in a synthetic graph. (a) shows the structure of the synthetic network. (b) shows the closeness centrality of the nodes, where in Y axis, nodes are sorted based on their centrality values. (c) shows the $R_u$ for all the nodes. Nodes in Y-axis of (c) are sorted in the same order as in (b). The colors of the lines in (b) and (c) correspond to three different types of nodes (colored accordingly) in (a). This figure also shows the high overlap between the top few nodes in both the lists.}
  \label{fig:visCentrality}
\end{figure}

As real life networks are noisy \cite{bandyopadhyay2019outlier}, first we experiment with a small synthetic graph as shown in Figure \ref{fig:visCentrality} to show the connection between $R_u$ and the centrality of the node $u \in V$. It has three communities and there is a central (red colored) node connecting all the communities. Each community has three sub-communities which are connected via the green colored nodes. The degree of each node in this network is kept roughly the same. We use closeness centrality \cite{opsahl2010node}, which is used widely in the network analysis literature. 
%Closeness centrality of a node $u \in V$ is defined as: $Closeness \ Centrality(u) = \frac{1}{\sum\limits_{v \in V, v \neq u} \; d_{uv}}$ where $d_{uv}$ = Shortest path-length between $u$ and $v$. This means if a node $u$ closer to rest of the nodes in a network, then it has a high closeness centrality. We create the synthetic graph as follows. Firstly, we synthesize a small network, call it a \textit{sub-community} (take any cluster of yellow nodes in Fig. \ref{fig:visCentrality}(a)) using stochastic block model\cite{SciPyProceedings_11} having intra-group probability of 0.5 and inter-group probability of 0.1. Next we add few nodes (green ones in Fig. \ref{fig:visCentrality}(a)) into this network and connect it with the small sub-communities of yellow vertices. We call this network as \textit{community}. We introduce a new node (red one in Fig. \ref{fig:visCentrality}(a)) and connect it with three copies of \textit{community}. The degree of each node in this network is kept roughly the same.
%
%The node colored with red in Fig \ref{fig:visCentrality} has the highest centrality in the graph, as it is well connected to all the communities and hence close to most of the nodes in the network. It is followed by the green color and then yellow color nodes in terms of centrality. 
The closeness centrality of the nodes are plotted in Fig. \ref{fig:visCentrality}(b). The nodes in the y-axis are sorted based on their closeness centrality values and as expected, the red node top the list as it is well connected to all the communities, followed by the green nodes, with yellow nodes placed at the bottom.
We run line2vec on this synthetic graph and plot the Radius $R_u$ for each node $u$ in Fig. \ref{fig:visCentrality}(c). Here also, the nodes are sorted in the same order as in sub-figure \ref{fig:visCentrality}(b). As one can see, the red node has the highest value of the radius. As this node is connected to a diverse set of nodes in the network, it needs a larger sphere to enclose the induced clique in the line graph. We also observe that most of the green nodes have higher values of $R_u$ than that of the yellow nodes. The correlation coefficient between the closeness centrality and the radius $R_u$ is 0.56. A more prominent trend can be observed for betweenness centrality \cite{skibski2016attachment}, where the correlation coefficient with the radius $R_u$ is 0.86.

On all the real-world datasets, we show the correlation of $R_u$ with the two centrality metrics for all the nodes in Table \ref{tab:corr}. High positive correlation between them can conclude that radius $R_u$ of a node is roughly proportional to the centrality of the node $u$ in the network. However, a detailed analysis is required to see the scope of introducing a new type of node centrality based on the values of $R_u$.

\begin{table}
    \caption{Pearson Correlation-Coefficient(CC) values obtained between the radius($R_u$) and centrality values of nodes for different networks. The centrality measures considered here are Betweenness and Closeness centrality.}
	\centering
% 	\scriptsize
    \resizebox{0.8\columnwidth}{!}{%
	\begin{tabular}{*6c}
	\toprule
	\sffamily{Dataset} & Karate & Dolphins & Cora & Pubmed & MSA \\
%    \sffamily{} & & & & Words & Distribution & links \\
    \hline
	\midrule
    \sffamily{Betweenness CC} & 0.81 & 0.66  & 0.29 & 0.26 & 0.35 \\
    \sffamily{Closeness CC} & 0.68 & 0.78 & 0.79 & 0.59 & 0.72 \\
\bottomrule
	\end{tabular}
    }
	%\vspace{-5mm}
	\label{tab:corr}
	\end{table} %sam modi 

\section{Discussion and Future Work}\label{sec:con}
We proposed a novel unsupervised dedicated edge embedding framework for homogeneous information and social networks.
We convert the given network to a weighted line graph and introduce the concept of collective homophily to embed the weighted line graph.
Our framework is quite generic. The skip-gram based component in the objective function of line2vec can easily be replaced with any other approach like graph convolution in weighted line graph.
Beside, we also plan to extend this methodology for heterogeneous information networks and knowledge bases.
There are several edge centric applications in networks. This work, being the first one towards a direct edge embedding, can play a basis to solve some of them in the context of network embedding and help to move network representation learning beyond node embedding.

%We propose a novel unsupervised approach called line2vec to generate embeddings of edges in information networks. Our solution maps the given network into a weighted line graph and then embed the line graph while exploiting the proposed concept of collective homophily. Experimental results reveal the benefit of such a direct edge embedding algorithm over the state-of-the-arts that embed edges by aggregating the node embeddings. Further analysis shows that line2vec generates a rich set of node embeddings as a by-product and is also related to the widely studied concept of node centrality in networks.
%
%There are several edge centric applications in networks. This work, being the first one towards a direct edge embedding, can play a basis to solve some of them in the context of network embedding. Also this is the perfect time where we can move network representation learning beyond node embedding and address fundamental tasks which are specific to network analysis.

\bibliography{ecai}
\end{document}

\title{Beyond Node Embedding: A Direct Unsupervised Edge Representation Framework via Collective Homophily for Information Networks}

% \author{Name1 Surname1 \and Name2 Surname2 \and Name3 Surname3\institute{University,
% Country, email: somename@university.edu} }

\maketitle
\bibliographystyle{ecai}

\begin{abstract}
This file contains additional materials to support the ideas presented in the main paper. Some of the materials present in this file can be incorporated into the camera ready version of the main paper, depending on the availability of the extra space and the feedback from the reviewers.
\end{abstract}

\section{lin2vec Algorithm}
Pseudo-code of line2vec algorithm is presented in Algorithm \ref{alg:.line2vec}. Referred sections are present in the main paper.

\begin{algorithm} %[H]%[tb]
  %\scriptsize
  \caption{\textbf{line2vec}}
  %\resizebox{0.47\textwidth}{!} {
  \label{alg:.line2vec}
%\resizebox{0.5\textwidth}{!}{\begin{minipage}{\textwidth}   
\begin{algorithmic}[1]
      
	\Statex \textbf{Input}: The network $G=(V,E)$, $K$: Dimension of the embedding space where $K<<m$
    \Statex \textbf{Output}: The edge embeddings of the network $G$
	\State Map the network $G$ to the weighted line graph $L(G) = (V_L,E_L)$, as described in Section 4.2 %\ref{sec:weightLG}.
	\State Use SGD to solve the optimization in Eq. 5 %\ref{eq:optiPen} 
	(with varying penalty parameters) as discussed in Section 4.3
	%\ref{sec:collecHom}.
	\State Get the embeddings $\mathbf{x}_{\mathbf{v}}$, $\mathbf{v} \in V_L$ of the nodes of the line graph, which is same as the embedding of the edges $(u,v)$ of the given network $G$, where $\mathbf{v} = l((u,v))$, as explained in Section 4.1 %\ref{sec:LT}.
	\end{algorithmic}
  \end{algorithm} %sam modi

\section{Complete Visualization Results}
\begin{figure*}[h!]
  \centering
  \begin{subfigure}[b]{0.15\linewidth}
    \includegraphics[width=\linewidth]{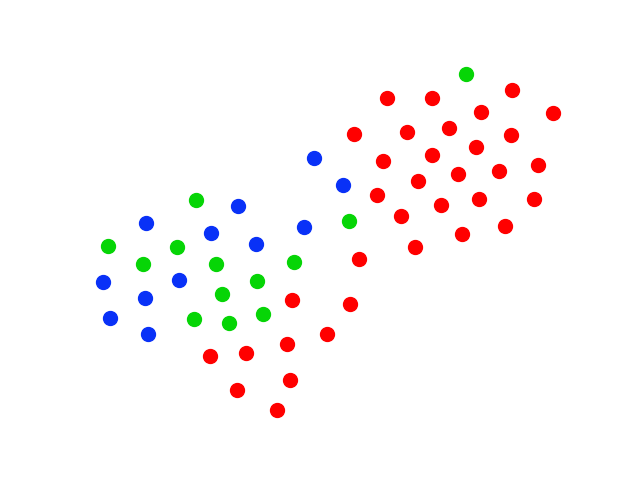}
    \caption{DeepWalk}
  \end{subfigure}
  \begin{subfigure}[b]{0.15\linewidth}
    \includegraphics[width=\linewidth]{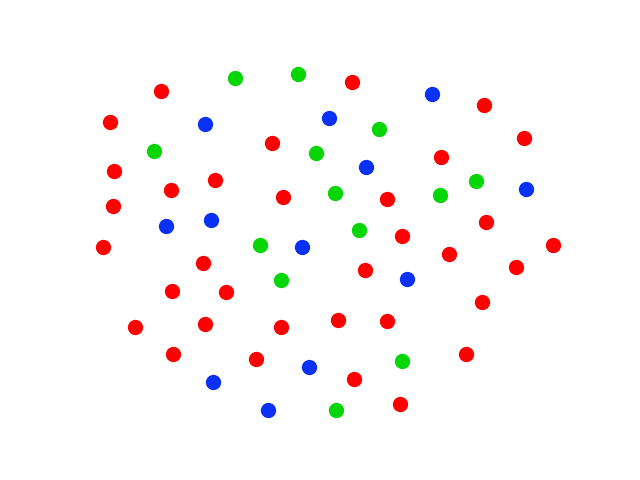}
    \caption{node2vec}
  \end{subfigure}
  \begin{subfigure}[b]{0.15\linewidth}
    \includegraphics[width=\linewidth]{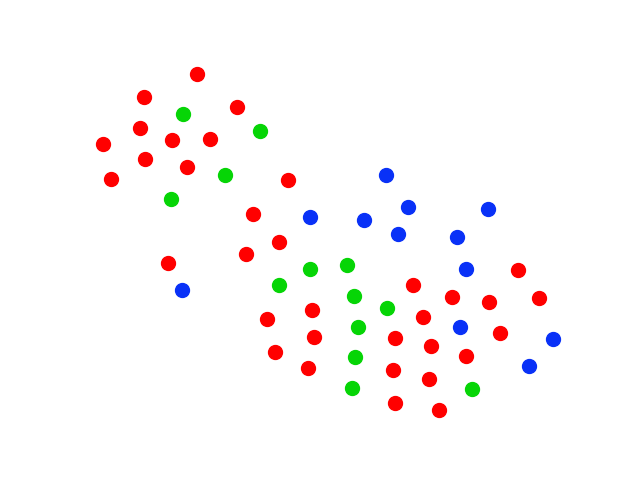}
    \caption{SDNE}
  \end{subfigure}
  \begin{subfigure}[b]{0.15\linewidth}
    \includegraphics[width=\linewidth]{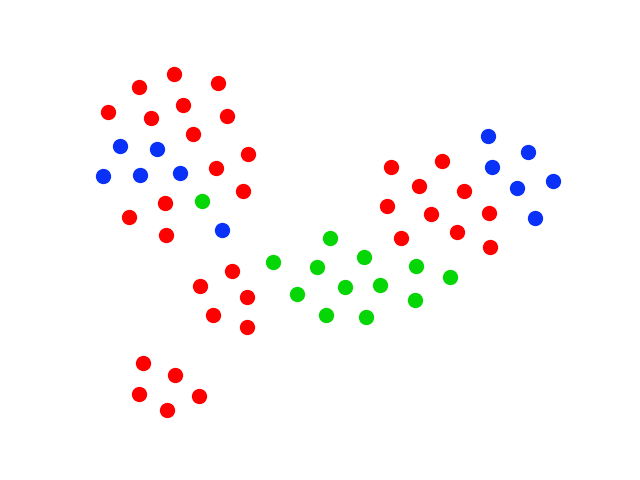}
    \caption{struc2vec}
  \end{subfigure}
  \begin{subfigure}[b]{0.15\linewidth}
    \includegraphics[width=\linewidth]{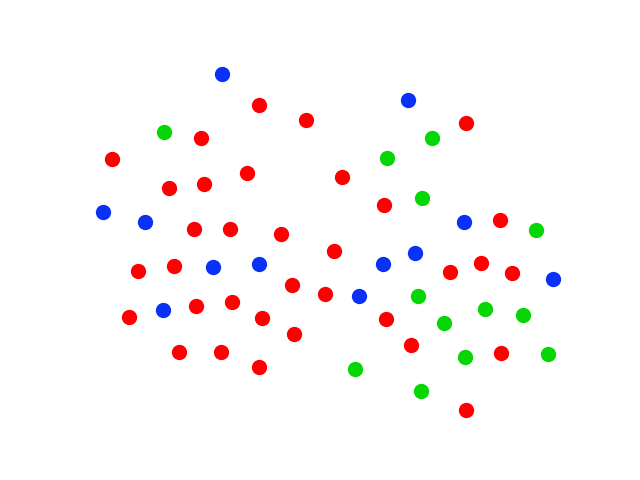}
    \caption{GraphSAGE}
  \end{subfigure}
  \begin{subfigure}[b]{0.15\linewidth}
    \includegraphics[width=\linewidth]{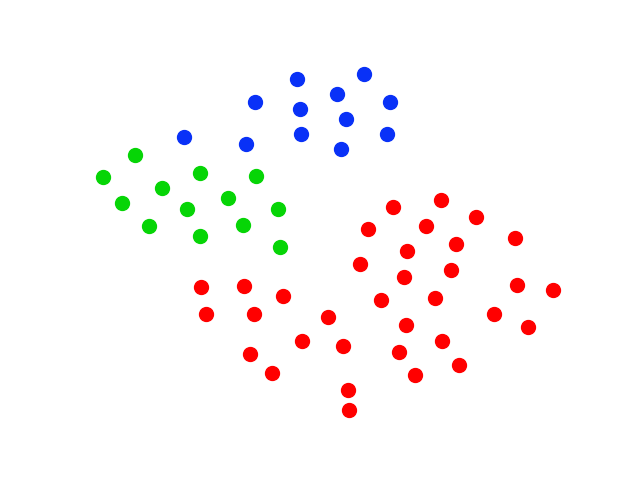}
    \caption{line2vec}
  \end{subfigure}
  \caption{Edge visualization on Karate dataset. Different colors represent different edge communities.}
  \label{fig:visKarate}
\end{figure*}

\begin{figure*}[h!]
  \centering
  \begin{subfigure}[b]{0.15\linewidth}
    \includegraphics[width=\linewidth]{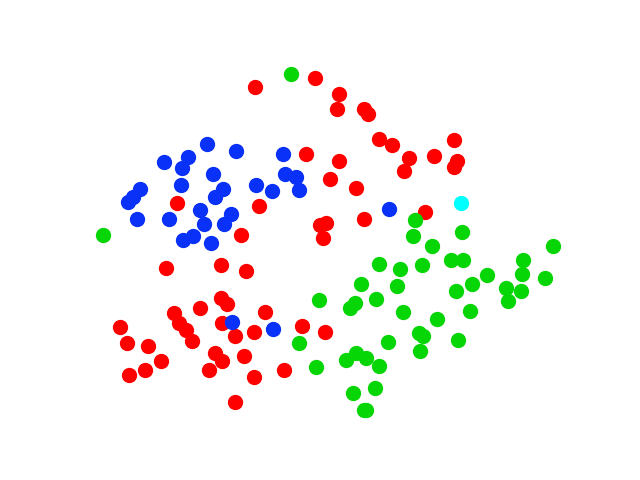}
    \caption{DeepWalk}
  \end{subfigure}
  \begin{subfigure}[b]{0.15\linewidth}
    \includegraphics[width=\linewidth]{images/visualization/dolphins/dolphins_n2v.png}
    \caption{node2vec}
  \end{subfigure}
  \begin{subfigure}[b]{0.15\linewidth}
    \includegraphics[width=\linewidth]{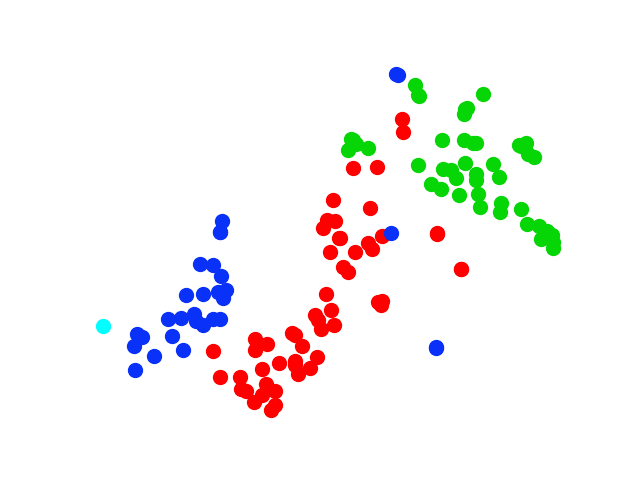}
    \caption{SDNE}
  \end{subfigure}
  \begin{subfigure}[b]{0.15\linewidth}
    \includegraphics[width=\linewidth]{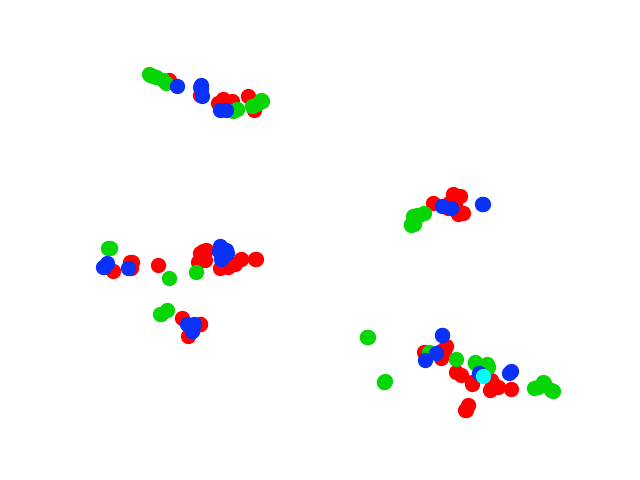}
    \caption{struc2vec}
  \end{subfigure}
  \begin{subfigure}[b]{0.15\linewidth}
    \includegraphics[width=\linewidth]{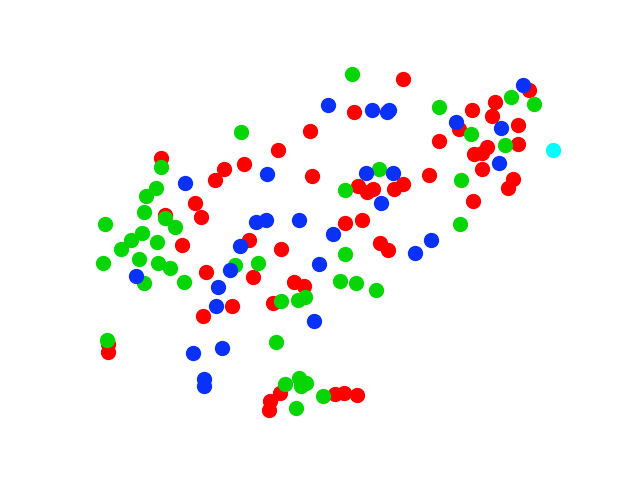}
    \caption{GraphSAGE}
  \end{subfigure}
  \begin{subfigure}[b]{0.15\linewidth}
    \includegraphics[width=\linewidth]{images/visualization/dolphins/dolphins_l2v.png}
    \caption{line2vec}
  \end{subfigure}
  \caption{Edge visualization on Dolphins dataset. Different colors represent different edge communities.}
  \label{fig:visDolphins}
\end{figure*}

\begin{figure*}[h!]
  \centering
  \begin{subfigure}[b]{0.16\linewidth}
    \includegraphics[width=\linewidth]{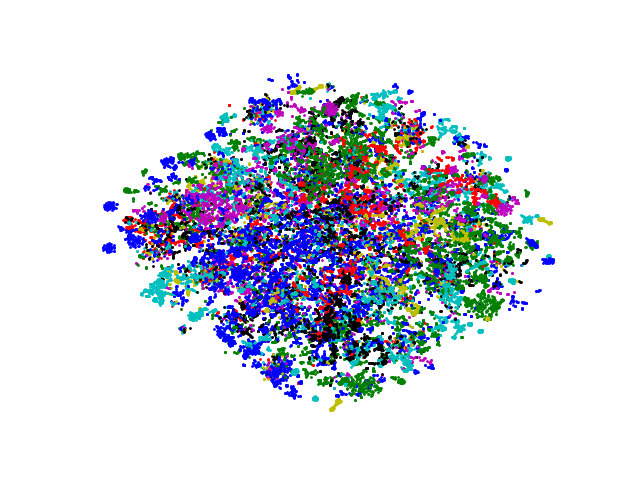}
    \caption{DeepWalk}
  \end{subfigure}
  \begin{subfigure}[b]{0.16\linewidth}
    \includegraphics[width=\linewidth]{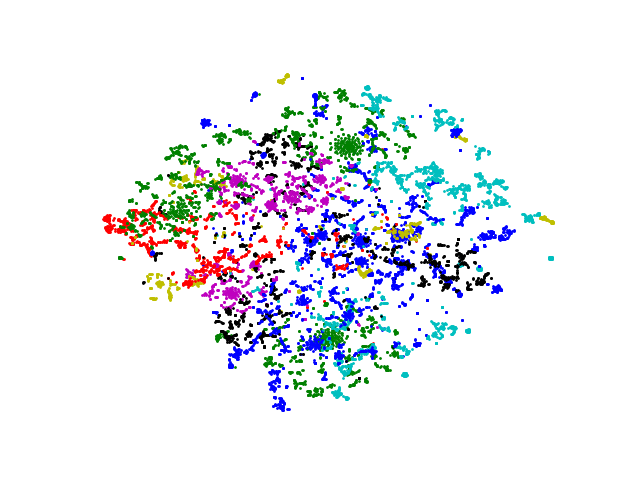}
    \caption{node2vec}
  \end{subfigure}
  \begin{subfigure}[b]{0.16\linewidth}
    \includegraphics[width=\linewidth]{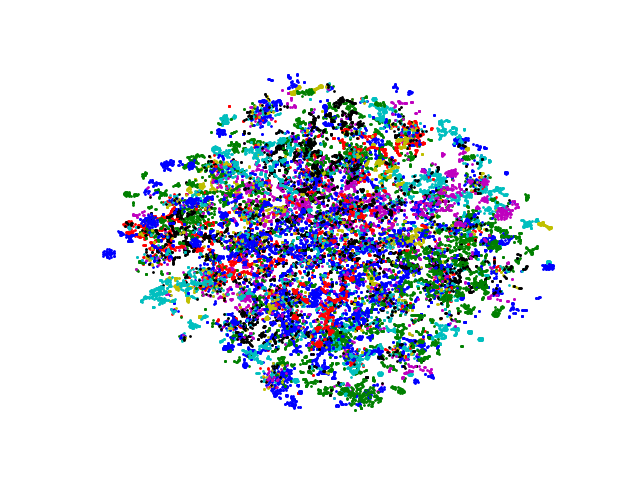}
    \caption{SDNE}
  \end{subfigure}
  \begin{subfigure}[b]{0.16\linewidth}
    \includegraphics[width=\linewidth]{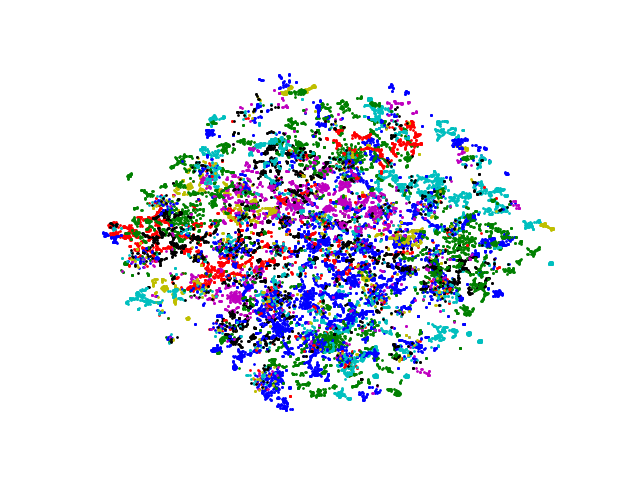}
    \caption{struc2vec}
  \end{subfigure}
  \begin{subfigure}[b]{0.16\linewidth}
    \includegraphics[width=\linewidth]{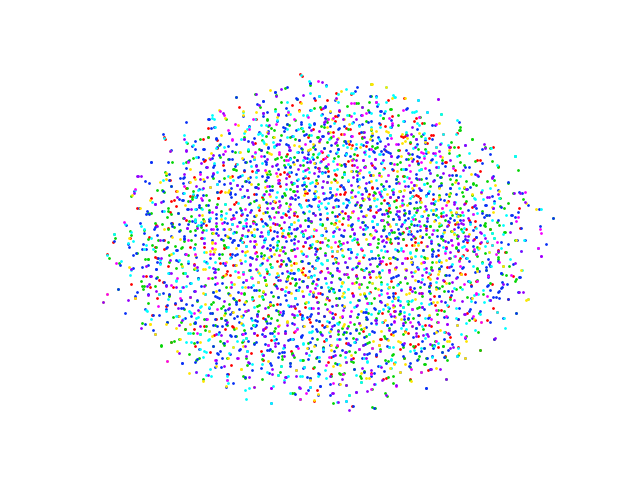}
    \caption{GraphSAGE}
  \end{subfigure}
  \begin{subfigure}[b]{0.16\linewidth}
    \includegraphics[width=\linewidth]{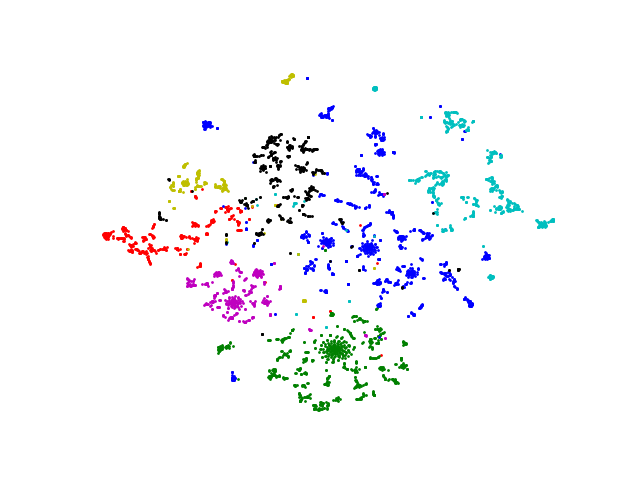}
    \caption{line2vec}
  \end{subfigure}
  \caption{Edge visualization on Cora dataset. Different colors represent different edge communities.}
  \label{fig:visCora}
\end{figure*}

\begin{figure*}[h!]
  \centering
  \begin{subfigure}[b]{0.16\linewidth}
    \includegraphics[width=\linewidth]{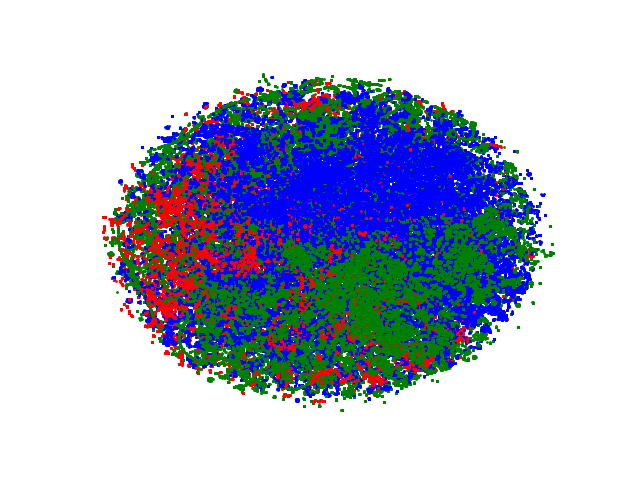}
    \caption{DeepWalk}
  \end{subfigure}
  \begin{subfigure}[b]{0.16\linewidth}
    \includegraphics[width=\linewidth]{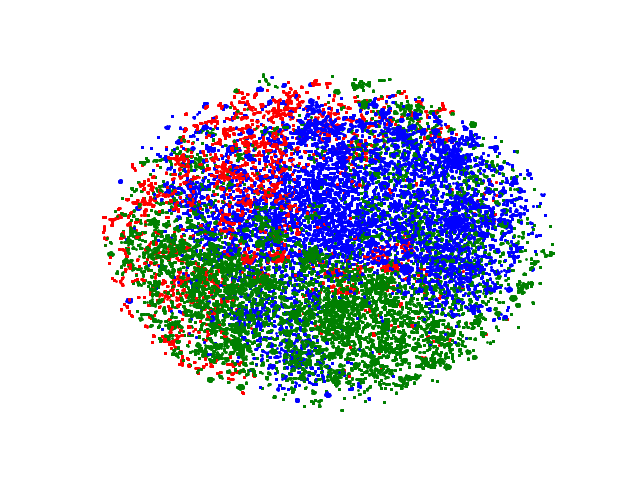}
    \caption{node2vec}
  \end{subfigure}
  \begin{subfigure}[b]{0.16\linewidth}
    \includegraphics[width=\linewidth]{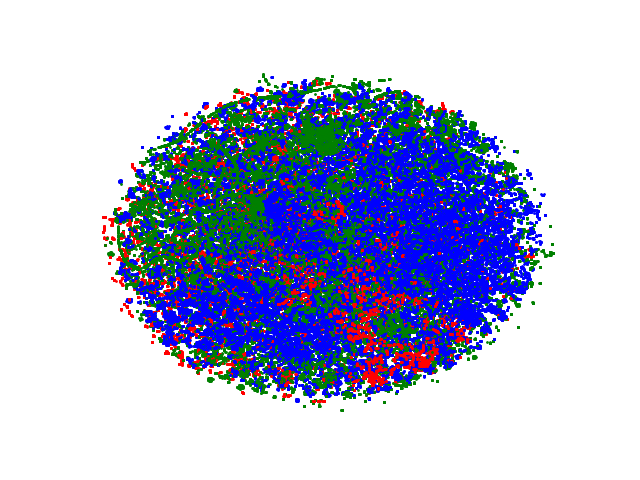}
    \caption{SDNE}
  \end{subfigure}
  \begin{subfigure}[b]{0.16\linewidth}
    \includegraphics[width=\linewidth]{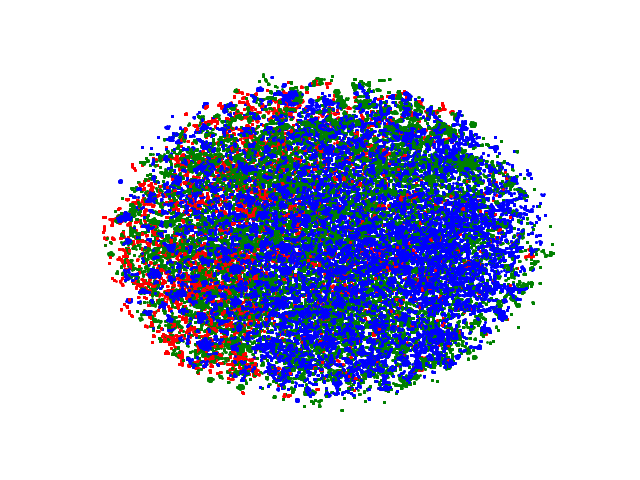}
    \caption{struc2vec}
  \end{subfigure}
  \begin{subfigure}[b]{0.16\linewidth}
    \includegraphics[width=\linewidth]{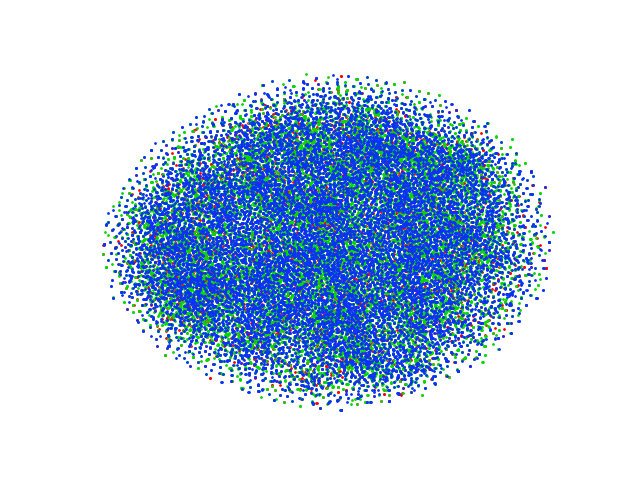}
    \caption{GraphSAGE}
  \end{subfigure}
  \begin{subfigure}[b]{0.16\linewidth}
    \includegraphics[width=\linewidth]{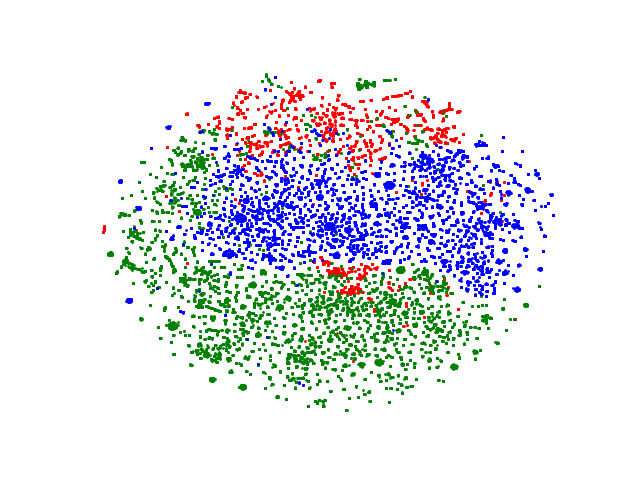}
    \caption{line2vec}
  \end{subfigure}
  \caption{Edge visualization on Pubmed dataset. Different colors represent different edge communities.}
  \label{fig:visPubmed}
\end{figure*}

\begin{figure*}[h!]
  \centering
  \begin{subfigure}[b]{0.16\linewidth}
    \includegraphics[width=\linewidth]{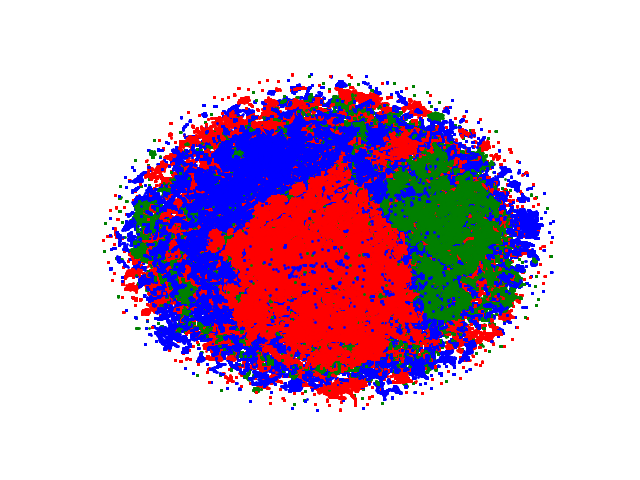}
    \caption{DeepWalk}
  \end{subfigure}
  \begin{subfigure}[b]{0.16\linewidth}
    \includegraphics[width=\linewidth]{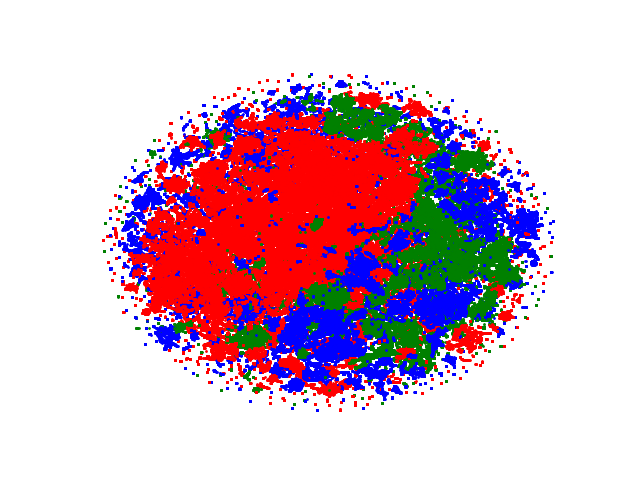}
    \caption{node2vec}
  \end{subfigure}
  \begin{subfigure}[b]{0.16\linewidth}
    \includegraphics[width=\linewidth]{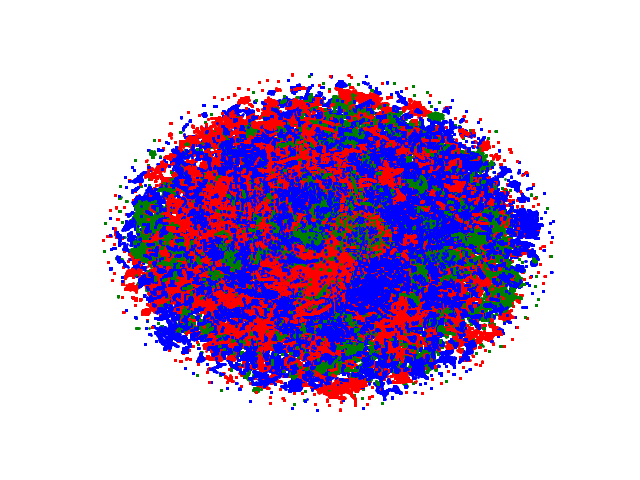}
    \caption{SDNE}
  \end{subfigure}
  \begin{subfigure}[b]{0.16\linewidth}
    \includegraphics[width=\linewidth]{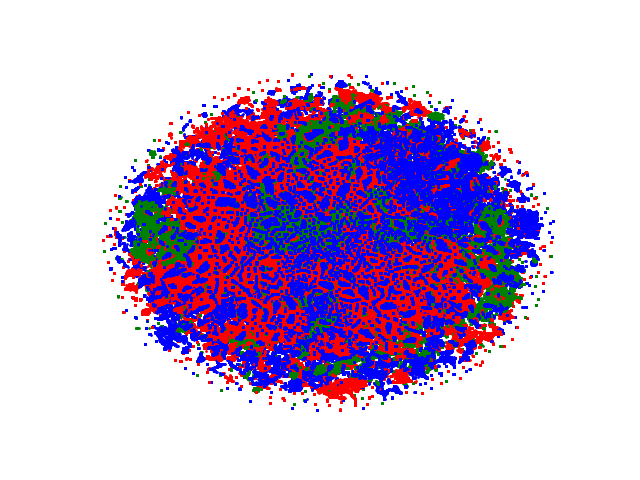}
    \caption{struc2vec}
  \end{subfigure}
  \begin{subfigure}[b]{0.16\linewidth}
    \includegraphics[width=\linewidth]{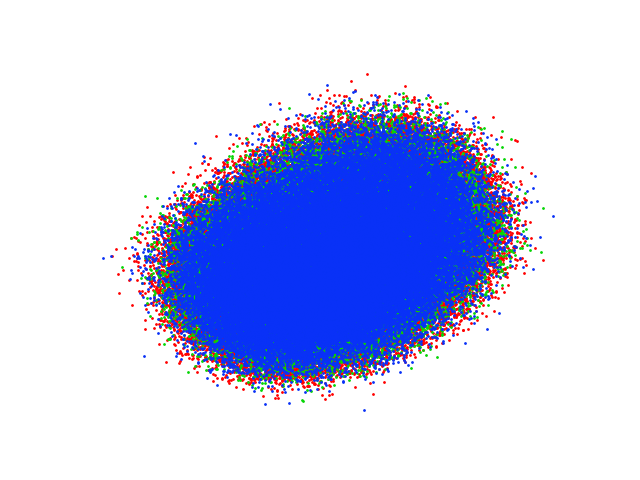}
    \caption{GraphSAGE}
  \end{subfigure}
  \begin{subfigure}[b]{0.16\linewidth}
    \includegraphics[width=\linewidth]{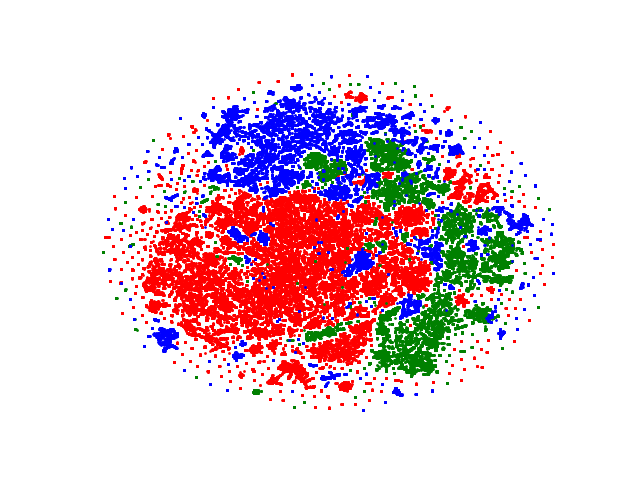}
    \caption{line2vec}
  \end{subfigure}
  \caption{Edge visualization on MSA dataset. Different colors represent different edge communities.}
  \label{fig:visMSA}
\end{figure*}

Please refer Figures \ref{fig:visKarate} to \ref{fig:visMSA} for the results of edge visualization on all the datasets.

\section{Complete set of Edge Classification Results}
\begin{figure}%[h]
\centering
\includegraphics[scale=0.22]{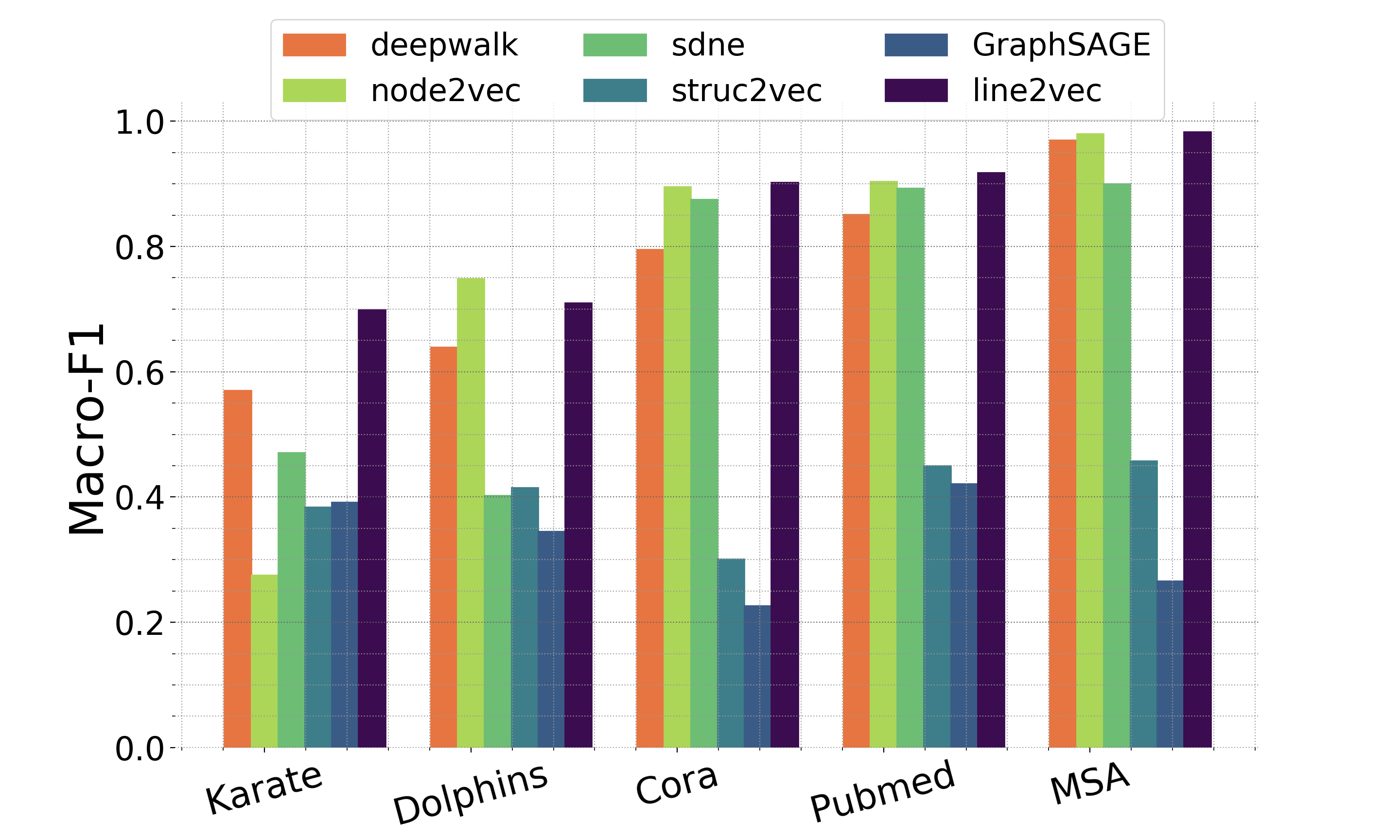}
\caption{Macro F1 Score of Edge Classification}
\label{fig:Macclassification}
\end{figure}

\begin{figure}%[h]
\centering
\includegraphics[scale=0.22]{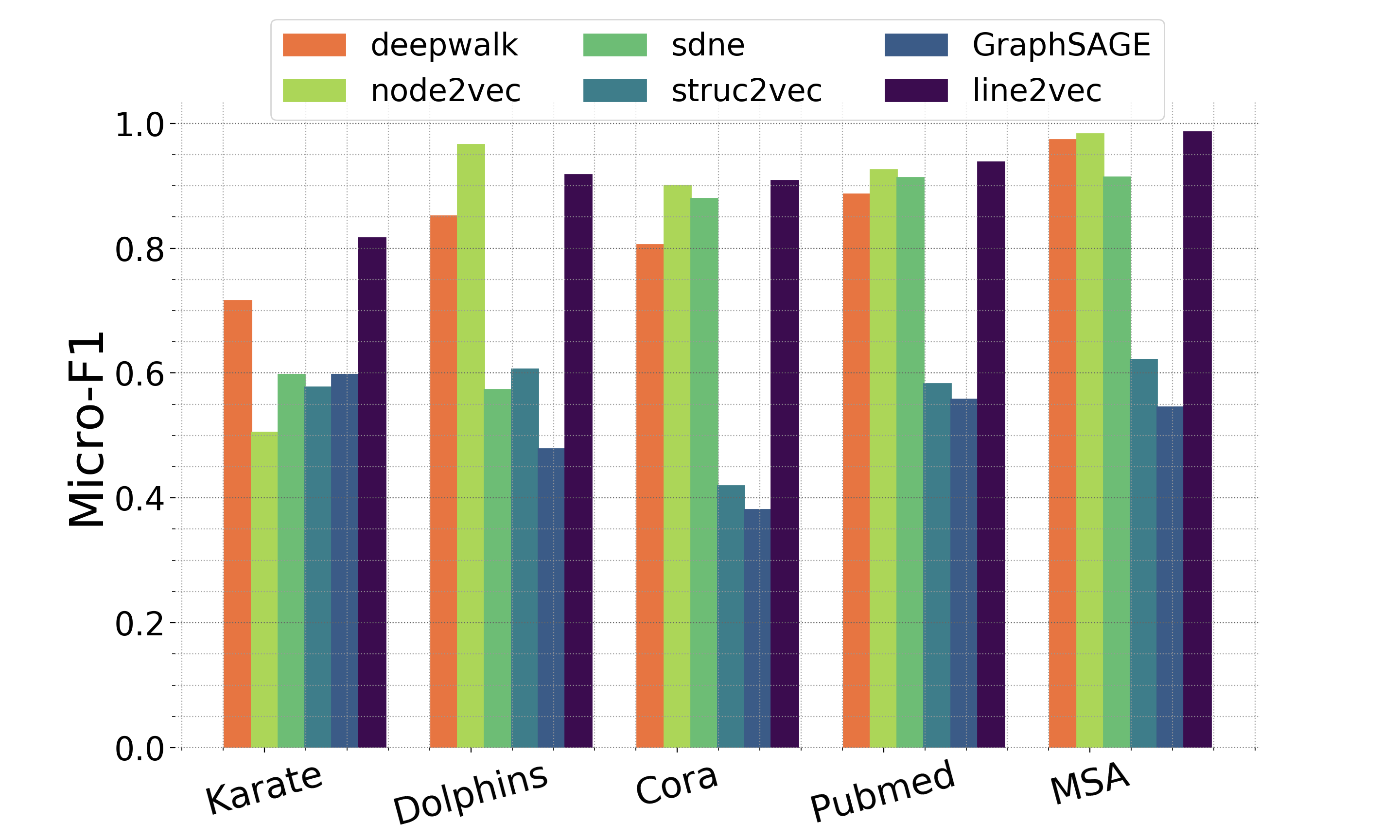}
\caption{Micro F1 Score of Edge Classification}
\label{fig:Micclassification}
\end{figure}

Figures \ref{fig:Macclassification} and \ref{fig:Micclassification} show the results of edge classification in terms of macro-F1 and micro-F1 scores.

\bibliography{ecai}